\documentclass{jfm}
\usepackage{graphicx}
\usepackage{epstopdf, epsfig}
\usepackage{mathrsfs}
\usepackage{amsmath}
\usepackage{amssymb}
\usepackage{color}
\usepackage[left]{lineno}

\shorttitle{The Structure of Turbulence in Unsteady Flow over Urban Canopies}

\title{The Structure of Turbulence in Unsteady Flow over Urban Canopies}
\shortauthor{W. Li and M. G. Giometto}

\author{Weiyi Li\aff{1}
 \and Marco G. Giometto\aff{1}
 \corresp{\email{mg3929@columbia.edu}}}

\affiliation{\aff{1}Department of Civil Engineering and Engineering Mechanics, Columbia University, New York, NY 10027}

\begin{document}

\maketitle

\begin{abstract}
The topology of turbulent coherent structures is known to regulate the transport of energy, mass, and momentum in the atmospheric boundary layer (ABL).
While previous research has primarily focused on characterizing the structure of turbulence in stationary ABL flows, real-world scenarios frequently deviate from stationarity, giving rise to nuanced and poorly understood changes in the turbulence geometry and associated transport mechanisms.
This study sheds light on this problem by examining topological changes in ABL turbulence induced by non-stationarity and their effects on momentum transport. 
Results from a large-eddy simulation of pulsatile open channel flow over an array of surface-mounted cuboids are examined. 
The analysis reveals that the flow pulsation triggers a phase-dependent shear rate, and the ejection-sweep pattern varies with the shear rate during the pulsatile cycle.
From a turbulence structure perspective, it is attributed to the changes in the geometry of hairpin vortices. 
An increase (decrease) in the shear rate intensifies (relaxes) these structures, leading to an increase (decrease) in the frequency of ejections and an amplification (reduction) of their percentage contribution to the total momentum flux. 
Furthermore, the size of the hairpin packets undergoes variations, which depend on the geometry of the constituting hairpin vortices, yet the packet inclination preserves its orientation throughout the pulsatile cycle. 
These observations reinforce the important role non-stationarity holds in shaping the structure of ABL turbulence and the momentum transport mechanisms it governs.

\end{abstract}

\section{Introduction}\label{sec:intro}
Coherent turbulent structures, also known as organized structures, control the exchange of energy, mass, and momentum between the earth's surface and the atmosphere, as well as within engineering systems. 
In wall-bounded flows, these structures have been shown to carry a substantial fraction of the mean shear stress \citep{lohou2000numerical,katul2006relative}, kinetic energy \citep{carper2004role,huang2009analysis,dong2020coherent}, and scalar fluxes \citep{li2011coherent,wang2014turbulent,li2019contrasts}. 
It hence comes as no surprise that substantial efforts have been devoted to their characterization across many fields.  
These structures are of practical relevance in applications relating to biosphere-atmosphere interaction \citep{raupach1986experiments,pan2014large}, air quality control \citep{michioka2014large}, urban climate \citep{christen2007coherent}, oceanography \citep{dishen2009}, and energy harvesting \citep{ali2017turbulence}, to name but a few.

Previous studies on coherent structures in atmospheric boundary layer (ABL) flows have mainly focused on the roughness sublayer (RSL) and the inertial sublayer (ISL)---the lower portions of the ABL. 
These layers host physical flow phenomena regulating land-atmosphere exchanges at scales relevant to weather models and human activities \citep{stull1988introduction, oke2017}. 
The RSL, which extends from the surface up to 2 to 5 times the average height of roughness elements, is characterized by flow heterogeneity due to the presence of these elements \citep{fernando2010fluid}. 
In the RSL, the geometry of turbulent structures is mainly determined by the underlying surface morphology. 
Through field measurements and wind tunnel data of ABL flow over vegetation canopies, \cite{raupach1996coherent} demonstrated that coherent structures near the top of a vegetation canopy are connected to inflection-point instabilities, akin to those found in mixing layers.
Expanding on the framework of this mixing-layer analogy, \cite{finnigan2009turbulence} employed conditional averaging techniques to show that the prevalent eddy structure in the RSL is a head-down hairpin vortex followed by a head-up one. 
This pattern is characterized by a local pressure peak and a strong scalar front located between the hairpin pair. 
More recently, \cite{bailey2016creation} challenged this observation by proposing an alternative two-dimensional roller structure with streamwise spacing that scales with the characteristic length suggested by \cite{raupach1996coherent}.

Extending the mixing-layer analogy to the urban RSL has proven challenging. 
In a numerical simulation study, \cite{coceal2007structure} discovered the absence of Kelvin-Helmholtz waves, which are a characteristic of the mixing-layer analogy, near the top of the urban canopy. 
This finding, corroborated by observations from \cite{huq2007shear}, suggests that the mixing-layer analogy is not applicable to urban canopy flows. 
Instead, the RSL of urban canopy flows is influenced by two length scales; the first is dictated by the size of individual roughness elements such as buildings and trees, and the second by the imprint of large-scale motions above the RSL. 
The coexistence of these two length scales can be observed through two-point correlation maps \citep{castro2006turbulence, reynolds2008measurements} and velocity spectra \citep{basley2019structure}. 
However, when the urban canopy has a significant aspect ratio between the building height $h$ and width $w$, such as $h/w > 4$, the momentum transport in the RSL is dominated by mixing-layer-type eddies, as shown by \cite{zhang2022evidence}.

The ISL, located above the RSL, is the geophysical equivalent of the celebrated law-of-the-wall region in high Reynolds number turbulent boundary layer (TBL) flows. 
In the absence of thermal stratification effects, mean flow in the ISL displays a logarithmic profile, and the momentum flux remains approximately constant with height \citep{stull1988introduction,marusic2013logarithmic,klewicki2014self}.
Surface morphology has been shown to impact ISL turbulence under certain flow conditions, and this remains a topic of active research.
\cite{volino2007turbulence} highlighted the similarity of coherent structures in the log region of TBL flow over smooth and three-dimensional rough surfaces via a comparison of velocity spectra and two-point correlations of the fluctuating velocity and swirl. 
Findings therein support Townsend's similarity hypothesis \citep{townsend1976structure}, which states that turbulence dynamics beyond the RSL do not depend on surface morphological features, except via their role in setting the length and velocity scales for the outer flow region.
The said structural similarity between TBL flows over different surfaces was later confirmed by \cite{wu2007outer} and \cite{coceal2007structure}, where a highly irregular rough surface and an urban-like roughness were considered, respectively.
However, \cite{volino2011turbulence} later reported pronounced signatures of surface roughness on flow structures beyond the RSL in a TBL flow over two-dimensional bars.
Similar observations were also made in a TBL flow over a surface characterized by cross-stream heterogeneity \citep{anderson2015numerical2}, thus questioning the validity of Townsend's similarity hypothesis.
To reconcile these contrasting observations, \cite{squire2017applicability} argued that structural similarity in the ISL is contingent on the surface roughness features not producing flow patterns significantly larger than their own size. 
If the surface-induced flow patterns are larger than their own size, then they may control flow coherence in the ISL.
For example, cross-stream heterogeneous rough surfaces can induce secondary circulations as large as the boundary-layer thickness, which profoundly modify momentum transport and flow coherence in the ISL \citep{barros2014observations,anderson2015numerical2}.

Although coherent structures in cases with significant surface-induced flow patterns necessitate case-specific analyses, researchers have extensively worked towards characterizing the topology of turbulence in cases that exhibit ISL structural similarity.
These analyses have inspired scaling laws \citep{meneveau2013generalized,yang2016moment,hu2023general} and the construction of statistical models \citep{perry1982mechanism} for TBL turbulence.
In this context, the hairpin vortex packet paradigm has emerged as the predominant geometrical model \citep{christensen2001statistical,tomkins2003spanwise,adrian2007hairpin}.
The origins of this model can be traced back to the pioneering work of \cite{theodorsen1952mechanisms}, who hypothesized that inclined hairpin or horseshoe-shaped vortices were the fundamental elements of TBL turbulence.
This idea was later supported by flow visualizations from laboratory experiments \citep{bandyopadhyay1980large,head1981new,smith1991dynamics} and high-fidelity numerical simulations \citep{moin1982numerical,moin1985structure,kim1986structure}.
In addition to providing evidence for the existence of hairpin vortices, \cite{head1981new} also proposed that these vortices occur in groups, with their heads describing an envelope inclined at $15^\circ – 20^\circ$ with respect to the wall. 
\cite{adrian2000vortex} expanded on this idea, and introduced the hairpin vortex packet paradigm, which posits that hairpin vortices are closely aligned in a quasi-streamwise direction, forming hairpin vortex packets with a characteristic inclination angle of $15^\circ – 20^\circ$. 
Nested between the legs of these hairpins are low-momentum regions, which extend approximately 2–3 times the boundary layer thickness in the streamwise direction. 
These low-momentum regions are typically referred to as large-scale motions \citep{smits2011high}. 
Flow visualization studies by \cite{hommema2003packet} and \cite{hutchins2012towards} further revealed that ABL structures in the ISL are also organized in a similar manner.

Of relevance for this work is that previous studies on coherent structures have predominantly focused on (quasi-)stationary flow conditions. 
However, stationarity is of rare occurrence in both ABL and engineering flow systems \citep{mahrt2020non, Lozano-duran2020}.
As discussed in the recent review paper by \cite{mahrt2020non}, there are two major drivers of non-stationarity in the ABL. 
The first involves temporal variations of surface heat flux, typically associated with evening transitions or the passage of individual clouds \citep{grimsdell2002observations}.
The second kind corresponds to time variations of the horizontal pressure gradient driving the flow, which can be induced by modes associated with propagating submeso-scale motions, mesoscale disturbances, and synoptic fronts \citep{monti2002observations,mahrt2014stably,cava2017wavelet}.
Previous studies have demonstrated that non-stationarity significantly affects flow statistics in the ABL, and can result in deviations from equilibrium turbulence
\cite{hicks2018relevance} reported that during morning and late afternoon transitions, the rapid change in surface heat flux disrupts the equilibrium turbulence relations. 
Additionally, several observational studies by Mahrt \textit{et al.} \citep{mahrt2007influence, mahrt2008influence, mahrt2013non} demonstrated that time variations in the driving pressure gradient can enhance momentum transport under stable atmospheric stratifications. 
Non-stationarity is also expected to impact the geometry of turbulence in the ABL, but this problem has not received much attention thus far. 
This study contributes to addressing this knowledge gap by investigating the impact of non-stationarity of the second kind on the topology of coherent structures in ABL turbulence and how it affects the mechanisms controlling momentum transport.
The study focuses on flow over urban-like roughness subjected to a time-varying pressure gradient. 
To represent flow unsteadiness, a pulsatile pressure gradient with a constant average and a sinusoidal oscillating component is selected as a prototype. 
In addition to its practical implications in areas such as wave-current boundary layers, internal-wave induced flows, and arterial blood flows, this flow regime facilitates the analysis of coherent structures, owing to the periodic nature of flow statistics.

Pulsatile flows share some similarities with oscillatory flows, i.e., flow driven by a time-periodic pressure gradient with zero mean. 
Interestingly, in the context of oscillatory flows, several studies have been devoted to the characterization of coherent structures. 
For instance, \cite{costamagna2003coherent, salon2007numerical} carried out a numerical study on transitional and fully turbulent oscillatory flow over smooth surfaces, and observed that streaky structures form at the end of the acceleration phases, then distort, intertwine, and eventually break into small vortices.
\cite{carstensen2010coherent} performed a series of laboratory experiments on transitional oscillatory flow, and identified two other major coherent structures, namely, cross-stream vortex tubes, which are the direct consequences of inflectional-point shear layer instability, and turbulent spots, which result from the destruction of near-wall streaky structures as those in stationary flows.
\cite{carstensen2012note} observed turbulent spots in oscillatory flows over sand-grain roughness, suggesting that the presence of such flow structures is independent of surface types, and it was later highlighted by \cite{mazzuoli2019turbulent} that the mechanism responsible for the turbulent spot generation is similar over both smooth and rough surfaces.
Although the primary modes of variability in oscillatory flows are relatively well understood, the same cannot be said for pulsatile flows. 
A notable study by \cite{zhang2019experimental} on wave-current boundary layers, a form of pulsatile flow, revealed phase variations in the spacing of streaks during the wave cycle. 
However, a detailed analysis of this particular behavior is still lacking.

To investigate the structure of turbulence in current-dominated pulsatile flow over surfaces in fully-rough aerodynamic flow regimes, we conduct a wall-modeled large-eddy simulation (LES) of flow over an array of surface-mounted cuboids. 
This study builds on the findings of a companion study that was recently accepted for publication in the Journal of Fluid Mechanics, focusing on the time evolution of flow statistics in pulsatile flow over a similar surface \citep{li2023mean}.
By contrasting findings against a corresponding stationary flow simulation, this study addresses these specific questions: (i) Does flow unsteadiness alter the topology of coherent structures in a time-averaged sense? (ii) How does the geometry of coherent structures evolve throughout the pulsation cycle? (iii) What is the effect of such modifications on the mechanisms governing momentum transfer in the ABL?
Answering these questions will achieve a twofold research objective: first, contributing to a better understanding of coherent patterns in pulsatile flow over complex geometries, and second, shedding light on how these patterns regulate momentum transfer.

This paper is organized as follows.
Section~\ref{sec:methods} outlines the numerical procedure and the simulation setup.
First- and second-order statistics are presented and discussed in \S\ref{sec:res-stats}.
Section~\ref{sec:res-qa} focuses on a quadrant analysis, whereas \S\ref{sec:res-Rii} and \S\ref{sec:res-inst} interpret the flow field in terms of two-point correlations and instantaneous flow behavior. 
Further insight on the time evolution of turbulence topology is proposed in \S\ref{sec:res-condavg} via conditional averaging.
Concluding remarks are given in \S\ref{sec:conclusion}.

\section{Methodology}\label{sec:methods}

\subsection{Numerical procedure}\label{sec:num}

Simulations are carried out via an in-house LES algorithm \citep{albertson1999natural, albertson1999surface, giometto2016spatial}. 
The LES algorithm solves the spatially-filtered momentum and mass conservation equations, namely,

\begin{eqnarray}
    \frac{\partial u_i}{\partial t} +  u_j (\frac{\partial u_i}{\partial x_j}-\frac{\partial u_j}{\partial x_i}) & = &  - \frac{1}{ \rho}\frac{\partial P}{ \partial  x_i} - \frac{\partial \tau_{ij}}{\partial x_j} - \frac{1}{\rho} \frac{\partial P_\infty}{ \partial  x_1}\delta_{i1}+F_i
    \label{eq:momentum} \\
    \frac{\partial u_i}{\partial x_i} & = & 0 
    \label{eq:continuity}
\end{eqnarray}
where $ (u_1,u_2,u_3 )$ represent the filtered velocities along the streamwise $x_1$, cross-stream $x_2$, and wall-normal $x_3$ directions, respectively. 
The rotational form of the convective term is used to ensure kinetic energy conservation in the discrete sense in the inviscid limit \citep{orszag1975numerical}.
$\tau_{ij}$ is the deviatoric part of the kinematic subgrid-scale (SGS) stress tensor, parameterized via the Lagrangian scale-dependent dynamic (LASD) Smagorinsky model \citep{bou2005scale}. 
The flow is assumed to be in the fully rough aerodynamic regime, and viscous stresses are not considered.
$P=p+\rho \frac{1}{3}\tau_{ii}+\rho \frac{1}{2} u_i u_i$ is a modified pressure, which accounts for the trace of SGS
stress and resolved turbulent kinetic energy, and $\rho$ is a constant fluid density.
The flow is driven by a spatially uniform, pulsatile pressure gradient in the $x_1$ direction, namely ${\partial P_\infty}/{ \partial  x_1} = -\rho f_m\left[ 1+\alpha_p \sin(\omega t) \right]$, where
the $f_m$ parameter controls the magnitude of the temporally averaged pressure gradient, $\alpha_p$ controls the forcing amplitude, and $\omega$ the forcing frequency. 
$\delta_{ij}$ in (\ref{eq:momentum}) denotes the Kronecker delta tensor.

Periodic boundary conditions apply in the wall-parallel directions, and a free-slip boundary condition is imposed at the top of the computational domain. 
The lower surface consists of an array of uniformly distributed cuboids, which are explicitly resolved via a discrete forcing immersed boundary method (IBM) \citep{mittal2005immersed}.  
The IBM approach makes use of an artificial force $F_i$ to impose the no-slip boundary condition at the solid-fluid interfaces.
Additionally, it utilizes an algebraic equilibrium wall-layer model to evaluate surface stresses \citep{piomelli2008wall, bose2018wall}.
The algorithm has been extensively validated for the simulation of flow in urban environments \citep[see, e.g.,][]{tseng2006modeling, chester2007modeling, giometto2016spatial}.

Spatial derivatives in the wall-parallel directions are computed via a pseudo-spectral collocation method based on truncated Fourier expansions \citep{orszag1970analytical}, whereas a second-order staggered finite differences scheme is employed in the wall-normal direction. 
Since dealiasing errors are known to be detrimental for pseudo-spectral discretization \citep{margairaz2018comparison}, non-linear convective terms are de-aliased exactly via the $3/2$ rule \citep{canuto2007spectral}. 
The time integration is performed via a second-order Adams-Bashforth scheme, and the incompressibility condition is enforced via a fraction step method \citep{kim1985application}.

\subsection{Simulation setup}\label{sec:cases}

\begin{figure}
  \centerline{\includegraphics[scale=0.4]{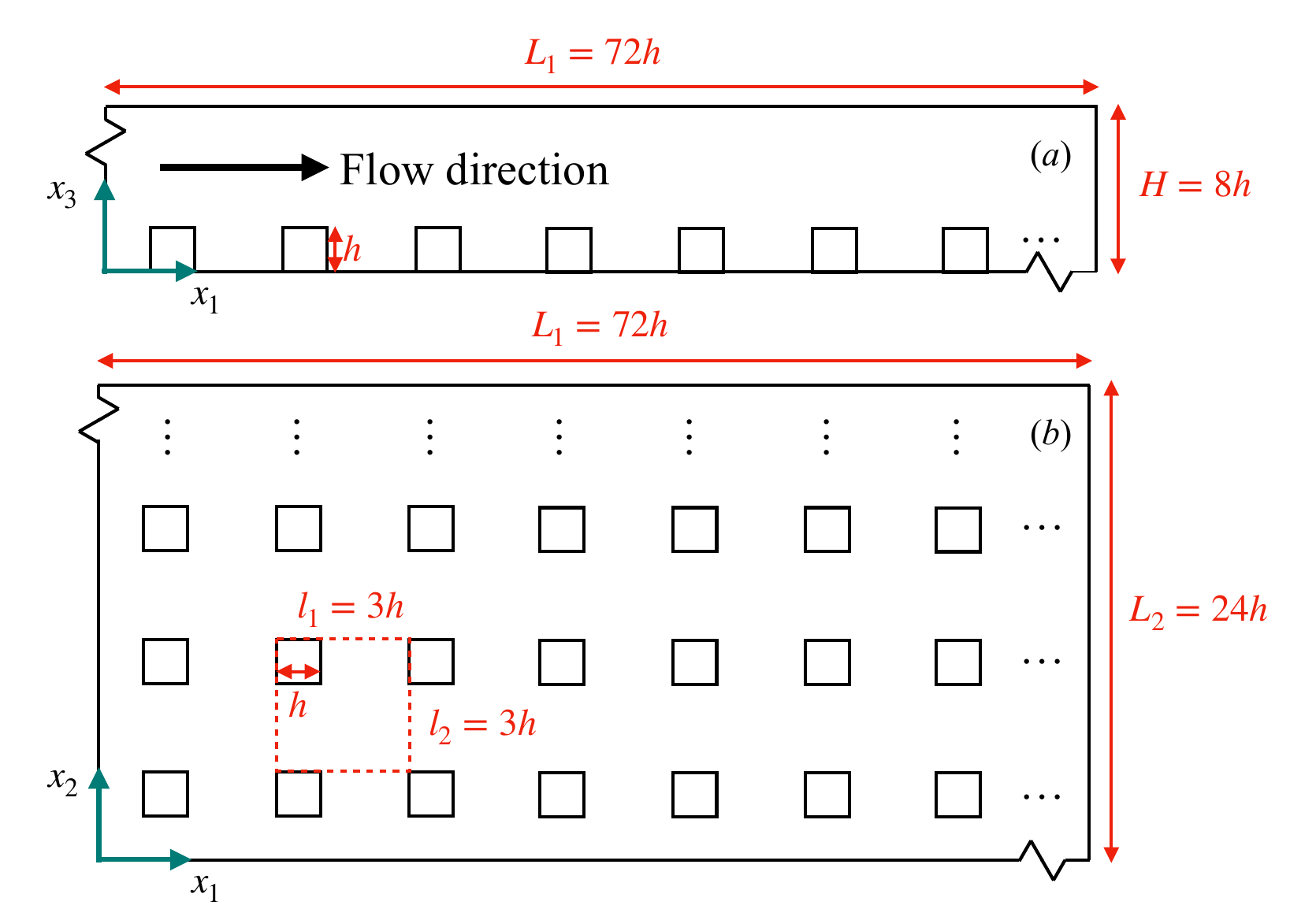}}
  \caption{Side and planar view of the computational domain (\textit{a},\textit{b} respectively). The red dashed line denotes the repeating unit.}
\label{fig:flow_config}

\end{figure}

Two LESs of flow over an array of surface-mounted cubes are carried out. 
The two simulations only differ by the pressure forcing term: One is characterized by a pressure gradient that is constant in space and time (CP hereafter), and the other by a pressure gradient that is constant in space and pulsatile in time (PP).

The computational domain for both simulations is sketched in figure \ref{fig:flow_config}.
The size of the box is $[0,L_1]\times[0,L_2]\times[0,H]$ with $L_1=72h$, $L_2=24h$ and $H=8h$, where $h$ denotes the height of cubes. 
Cubes are organized in an in-line arrangement with planar and frontal area fractions set to $\lambda_p = \lambda_f = 0.\overline{1}$.  
The relatively high packing density along with the chosen scale separation $H/h = 8$ support the existence of a well-developed ISL and healthy coherent structures in the considered flow system \citep{coceal2007structure,castro2007rough,zhang2022evidence}. 
In terms of horizontal extent, $L_1/H$ and $L_2/H$ are larger than those from previous works focusing on coherent structures above aerodynamically rough surfaces \citep{coceal2007structure, xie2008large, leonardi2010channel, anderson2015numerical} and are sufficient to accommodate large-scale motions \citep{balakumar2007large}. 
An aerodynamic roughness length $z_0=10^{-4} h$ is prescribed at the cube surfaces and the ground via the algebraic wall-layer model, resulting in negligible SGS drag contributions to the total surface drag \citep{yang2016recycling}. 
The computational domain is discretized using a uniform Cartesian grid of $N_1 \times N_2 \times N_3 = 576 \times 192 \times 128$, so each cube is resolved via $8 \times 8 \times 16$ grid points. 
Such a grid resolution yields flow statistics that are poorly sensitive to grid resolution in both statistically stationary and pulsatile flows at the considered oscillation frequency \citep{tseng2006modeling,li2023mean}. 

For the PP case, the forcing frequency is set to $\omega T_h=\pi/8$, where $T_h = h/u_\tau$ is the averaged turnover time of characteristic eddies of the urban canopy layer (UCL) and ${u}_{\tau}=\sqrt{f_m H}$ the friction velocity. 
This frequency selection is based on both practical and theoretical considerations.
Realistic ranges for the friction velocity and UCL height are $0.1 \le {u}_{\tau} \le 0.5 \ \rm{m/s}$ and $3 \le h \le 30 \ \rm{m}$ \citep{stull1988introduction}.
Using such values, the chosen frequency corresponds to a time period $ 24 \le T \le 4800 \ \rm{s}$, where $T=2\pi/\omega = 16 T_h$.
This range of time scales pertains to sub-mesoscale motions  \citep{mahrt2009characteristics,hoover2015submeso}, which, as outlined in \S\ref{sec:intro}, are a major driver of atmospheric pressure gradient variability. 
From a theoretical perspective, this frequency is expected to yield substantial modifications of coherent structures within the ISL.
The chosen frequency results in a Stokes layer thickness $\delta_s=5h$, encompassing both the RSL and the ISL. 
Within the Stokes layer, turbulence generation and momentum transport undergo significant modifications during the pulsation cycle, as demonstrated in \cite{li2023mean}. 
Moreover, the oscillation period $T$ is comparable to the average lifespan of eddies in the ISL of the considered flow system, as elaborated below. 
\cite{coceal2007structure} showed that, in flow over rough surfaces, the characteristic length scale of ISL eddies ($\ell$) is bounded below by $h$, thus yielding $\min{(\ell)} \sim h$. 
Based on Townsend's attached-eddy hypothesis, $\ell \sim x_3$, which results in $\max{(\ell)} \sim H$. 
The time scale associated with ISL eddies is $T_\ell = \ell/u_\tau$, so that $\min{(T_\ell)}  \sim h/u_\tau = T_h$ and $\max{(T_\ell)} \sim H/u_\tau = T_H$.
The modest scale separation characterizing our setup ($H = 8h$) yields a modest separation of time scales in the ISL, and considering $T \approx T_H$, one can conclude that the time scale of ISL eddies is comparable to $T$.
With $T_\ell \approx T$, flow pulsation will considerably modify the structure of ISL turbulence and drive the flow out of equilibrium conditions. This is because changes in the imposed pressure gradient occur at a rate that enables turbulent eddies to respond.
This behavior can be contrasted to two limiting cases: with $T_\ell \gg T$, turbulence is unable to respond to the rapid changes in the environment and is advected in a ``frozen'' state, i.e., it does not undergo topological changes. 
With $T_\ell \ll T$, ISL eddies do not ``live'' long enough to sense changes in the environment, and maintain a quasi-steady state throughout the pulsatile cycle.
In terms of forcing amplitude, such a quantity is set to $\alpha_p=12$ for the PP case; this amplitude is large enough to induce significant changes in the coherent structures with the varying pressure gradient while avoiding mean flow reversals.

Both simulations are initialized with velocity fields from a stationary flow case and integrated over $400 T_{H}$, corresponding to $200$ pulsatile cycles for the PP case.
Here $T_{H}= H/{u}_{\tau}$ refers to the turnover time of the largest eddies in the domain.
The time step ($\delta t$) is set to ensure $\max{(CFL)}=u_{max} \delta t/\delta  \approx 0.05$, where CFL denotes the Courant-Friedrichs-Lewy stability condition, $u_{max}$ is the maximum velocity magnitude at any point in the domain during the simulation, and $\delta$ is the smallest grid stencil in the three coordinate directions.
The initial $20 T_{H}$ are discarded for both the CP and PP cases (transient period for the PP case), which correspond to about 10 oscillation periods, after which instantaneous snapshots of velocities and pressure are saved to disk every $0.025T_{H}$ ($1/80$ of the pulsatile cycle). 

\subsection{Notation and terminology}\label{sec:notation}
For the PP case, $\overline{(\cdot)}$ denotes an ensemble averaging operation, performed over the phase dimension and over repeating surface units (see figure \ref{fig:flow_config}), i.e., 
\begin{multline}
 \overline{\theta} (x_1,x_2,x_3,t)= \frac{1}{N_p n_1 n_2}\sum^{N_p}_{n=1} \sum^{n_1}_{i=1}\sum^{n_2}_{j=1} \theta(x_1+il_1,x_2+jl_2,x_3,t+nT) ,\\
  0\leq x_1 \leq l_1,\quad 0\leq x_2 \leq l_2,\quad 0\leq t \leq T \ ,
\label{eq:phavg}
\end{multline}
where $\theta$ is a given scalar field, $n_1$ and $n_2$ are the number of repeating units in the streamwise and cross-stream directions, respectively.
Using the usual Reynolds decomposition, one can write 
\begin{equation}
 \theta (x_1,x_2,x_3,t)=  \overline{\theta} (x_1,x_2,x_3,t)+\theta^\prime(x_1,x_2,x_3,t) \ \,
	\label{eq:phavg&prime}
\end{equation}
where $(\cdot)^\prime$ denotes a fluctuation from the ensemble average. 
For the CP case, $\overline{(\cdot)}$ denotes a quantity averaged over time and repeating units. 
An ensemble averaged quantity can be further decomposed into an intrinsic spatial average and a deviation from the intrinsic average \citep{schmid2019volume}, i.e., 
\begin{equation}
  \overline{\theta} (x_1,x_2,x_3,t)= \langle \overline{\theta} \rangle (x_3,t)+ \overline{\theta} ^{\prime \prime}(x_1,x_2,x_3,t) \ .
	\label{eq:intrinsic_avg}
\end{equation}
Note that, for each $x_3$, the intrinsic averaging operation is taken over a thin horizontal ``slab'' $V_f$ of fluid, characterized by a thickness $\delta_3$ in the wall-normal ($x_3$) direction, namely,
\begin{equation}
\langle \overline{\theta} \rangle (x_3,t)=  \frac{1}{V_f}\int_{x_3-\delta_3/2}^{x_3+\delta_3/2}\int_0^{l_2} \int_0^{l_1}\overline{\theta} (x_1,x_2,x_3,t) dx_1 dx_2 dx_3 \ .
	\label{eq:volavg}
\end{equation}

Further, any phase-averaged quantity from the PP case consists of a longtime-averaged component and an oscillatory component with a zero mean, which will be hereafter denoted via the subscripts $l$ and $o$, respectively, i.e.,
\begin{equation}
  \overline{\theta} (x_1,x_2,x_3,t)= \overline{\theta}_{l} (x_1,x_2,x_3)+  \overline{\theta}_{o} (x_1,x_2,x_3,t)
	\label{eq:bar_longtime_osc}
\end{equation}
and 
\begin{equation}
  \langle \overline{\theta} \rangle (x_3,t)= \langle \overline{\theta} \rangle_{l} (x_3) +  \langle \overline{\theta} \rangle_{o} (x_3,t)\ .
	\label{eq:bracket_longtime_osc}
\end{equation}

As for the CP case, the longtime and ensemble averages are used interchangeably due to the lack of an oscillatory component.
In the following, the longtime-averaged quantities from the PP case are contrasted against their counterparts from the CP case to highlight the impact of flow unsteadiness on flow characteristics in a longtime average sense. 
Oscillatory and phase-averaged quantities are analyzed to shed light on the phase-dependent features of the PP case.

\section{Results}

\subsection{Overview of flow statistics}\label{sec:res-stats}
\cite{li2023mean} have proposed a detailed analysis of pulsatile flow over an array of surface-mounted cuboids, discussing the impact of varying forcing amplitude and frequency on selected flow statistics. 
Here, we repropose and expand upon some of the findings for the chosen oscillation frequency and amplitude that are relevant to this work.

\begin{figure}
  \centerline{\includegraphics[width=\textwidth]{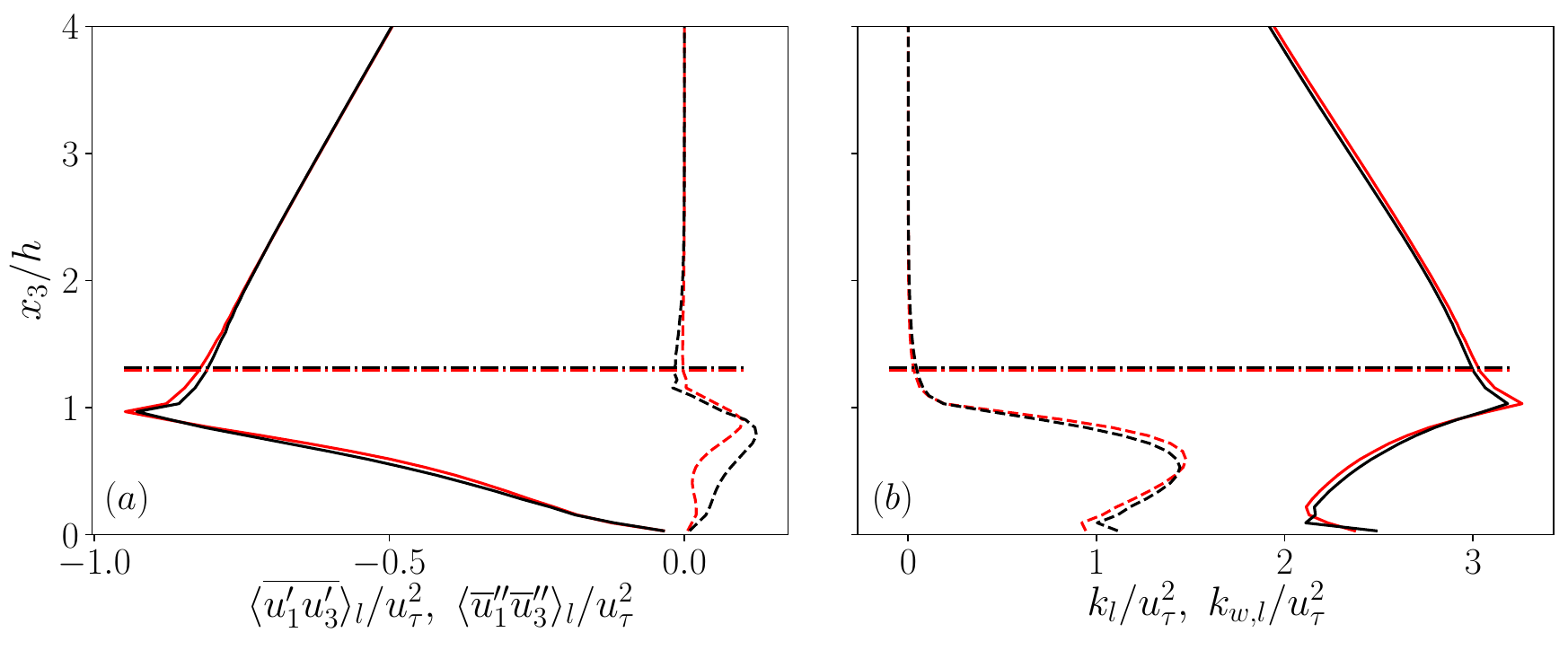}}
  \caption{(\textit{a}) Longtime-averaged shear stresses from the PP (black) and CP (red) cases. Resolved Reynolds shear stress $\langle \overline{u^\prime_1 u^\prime_3} \rangle_l$, solid lines; dispersive shear stress $\langle \overline{u}^{\prime \prime}_1 \overline{u}^{\prime \prime}_3 \rangle_l$. (\textit{b}) Longtime-averaged turbulent and wake kinetic energy from the PP (black) and CP (red) cases. Resolved turbulent kinetic energy $k_l = \langle \overline{u^\prime_i u^\prime_i} \rangle_l/2$, solid lines; wake kinetic energy $k_{w,l}=\langle \overline{u}^{\prime \prime}_i \overline{u}^{\prime \prime}_i \rangle_l/2$, dashed lines. Dashed-dotted horizontal lines denote the upper bound of the  RSL $(x_3^R)$. }
\label{fig:longtime_rey}
\end{figure}

Figure \ref{fig:longtime_rey}(\textit{a}) presents the wall-normal distributions of the longtime-averaged resolved Reynolds shear stress $\langle \overline{u^\prime_1 u^\prime_3} \rangle_l$ and dispersive shear stress $\langle \overline{u}^{\prime \prime}_1 \overline{u}^{\prime \prime}_3 \rangle_l$.
Note that SGS components contribute less than $1\%$ to the total Reynolds stresses and are hence not discussed.
From the figure, it is apparent that flow unsteadiness does not noticeably affect the $\langle \overline{u^\prime_1 u^\prime_3} \rangle_l$ profile, with local variations from the statistically stationary scenario being within a $3 \%$ margin. 
On the contrary, flow pulsation within the UCL leads to pronounced increases in $\langle \overline{u}^{\prime \prime}_1 \overline{u}^{\prime \prime}_3 \rangle_l$, with local surges reaching up to a fivefold increase. 
However, despite this increase, the dispersive flux remains a modest contributor to the total momentum flux in the UCL.
Figure \ref{fig:longtime_rey}(\textit{b}) displays the longtime-averaged resolved turbulent kinetic energy $k_l = \langle \overline{u^\prime_i u^\prime_i} \rangle_l/2$ and wake kinetic energy $k_{w,l} = \langle \overline{u}^{\prime \prime}_i \overline{u}^{\prime \prime}_i \rangle_l/2$.
Both $k_l$ and $k_{w,l}$ from the PP case feature modest ($<5\%$) local departures from their CP counterparts, highlighting a weak dependence of both longtime-averaged turbulent and wake kinetic energy on flow unsteadiness.
Also, the RSL thicknesses $(x_3^R)$ for the CP and PP cases are depicted in figure \ref{fig:longtime_rey}.
Following the approach by \cite{pokrajac2007quadrant}, $x_3^R$ is estimated by thresholding the spatial standard deviation of the longtime-averaged streamwise velocity normalized by its intrinsic average, namely,
\begin{equation}
 \sigma=\frac{\sqrt{\langle(\overline{u}_{1,l}-\langle\overline{u}_1\rangle_l)^2\rangle}}{\langle\overline{u}_1\rangle_l}\ ,
	\label{eq:rsl}
\end{equation}
where the threshold is taken as 1\%. 
An alternative method to evaluate $x_3^R$ involves using phase-averaged statistics instead of longtime-averaged ones in (\ref{eq:rsl}). 
Although not shown, such a method yields similar predictions (with a discrepancy of less than $5\%$).
Both $\langle \overline{u}^{\prime \prime}_1 \overline{u}^{\prime \prime}_3 \rangle_l$ and $k_{w,l}$ reduce to less than $1\%$ of their peak value above $x_3^R$.
From figure \ref{fig:longtime_rey}, one can readily observe that flow unsteadiness yields a modest increase in the extent of the RSL, with an estimated $x_3^R$ not exceeding $1.5h$ in both cases. 
Hereafter, we will hence assume $x_3^R=1.5h$.
As discussed in \S\ref{sec:intro}, RSL and ISL feature distinct coherent structures.
Specifically, the structures in the RSL are expected to show strong imprints of roughness elements, whereas those in the ISL should, in principle, be independent of surface morphology \citep{coceal2007structure}.

\begin{figure}
  \centerline{\includegraphics[width=\textwidth]{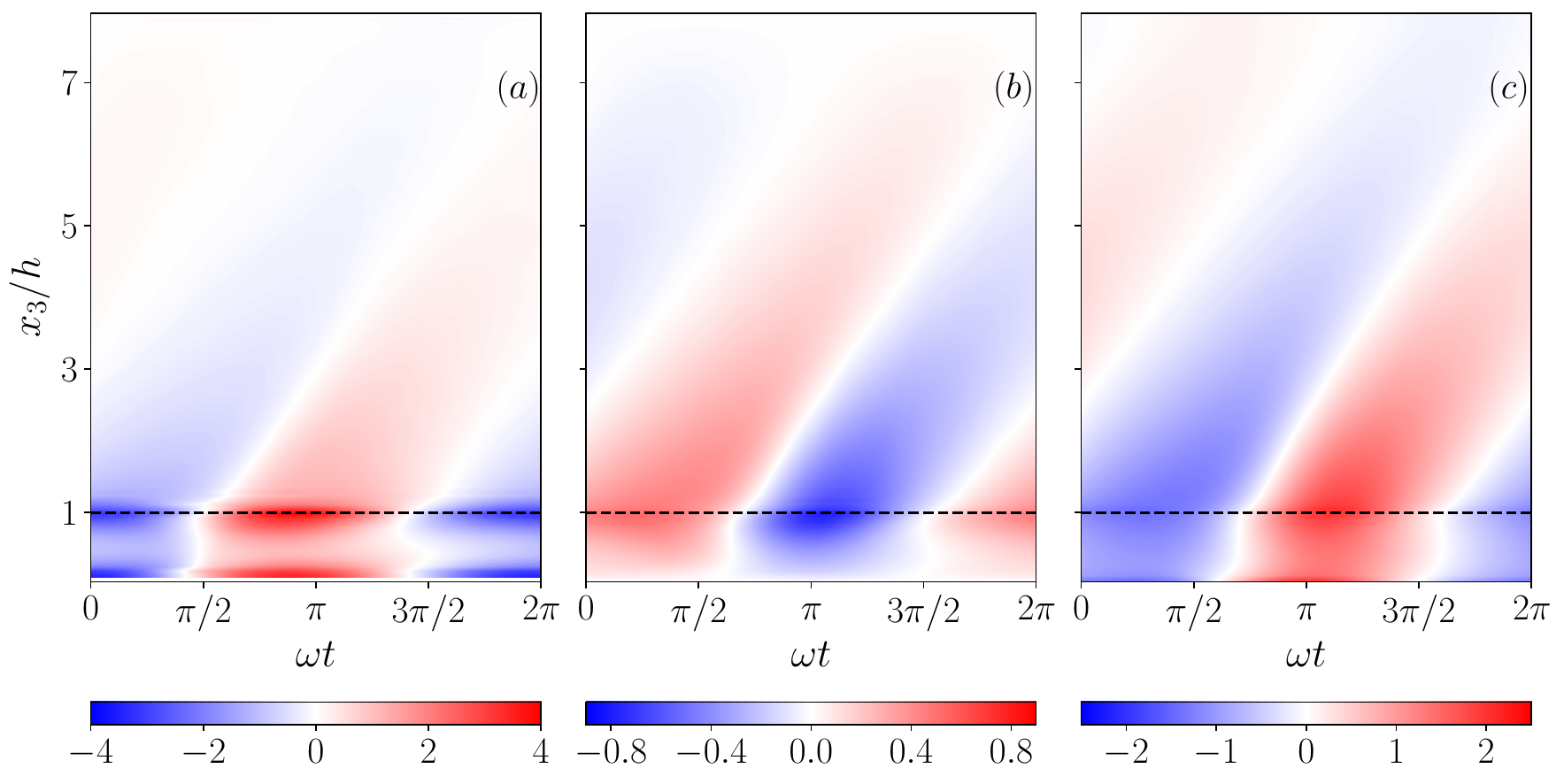}}
  \caption{Space-time diagrams of (\textit{a}) oscillatory shear rate ${\partial \langle \overline{u}_1 \rangle_o}/{\partial x_3}$, (\textit{b}) oscillatory resolved Reynolds shear stress $\langle \overline{u_1^\prime u_3^\prime}\rangle_o $ and (\textit{c}) oscillatory resolved turbulent kinetic energy $k_o=\langle \overline{u^\prime_i u^\prime_i}\rangle_o/2$ from the PP case. Results are normalized by $u_{\tau}$ and $h$. Horizontal dashed lines highlight the top of the UCL.}
\label{fig:cs_osc_temp_comp_paper}
\end{figure}

The response of selected first- and second-order flow statistics to flow unsteadiness is depicted in figure \ref{fig:cs_osc_temp_comp_paper}. 
In Figure \ref{fig:cs_osc_temp_comp_paper}(\textit{a}), an oscillating wave is evident in the oscillatory shear rate $\partial \langle \overline{u}_1 \rangle_o /\partial x_3 $. 
This wave, generated at the canopy top due to flow unsteadiness, exhibits a phase lag of $\pi/2$ relative to the pulsatile pressure forcing.
Such a wave propagates in the positive vertical direction while being attenuated and diffused by turbulent mixing. 
It is noteworthy that the propagation speed of the oscillating shear rate is to a good degree constant, as suggested by the constant tilting angle along the $x_3$ direction of the ${\partial \langle \overline{u}_1 \rangle_o}/{\partial x_3}$ contours.
As apparent from figure \ref{fig:cs_osc_temp_comp_paper}(\textit{b},\textit{c}), the space-time diagrams of the oscillatory resolved Reynolds shear stress $\langle \overline{u_1^\prime u_3^\prime}\rangle_o $ and oscillatory resolved turbulent kinetic energy $k_o=\langle \overline{u^\prime_i u^\prime_i}\rangle_o/2$ are also characterized by decaying waves traveling away from the RSL at constant rates.
The speeds of these waves are similar to that of the corresponding oscillating shear rate, which can be again inferred by the identical tilting angles in the contours. 
There is clearly a causal relation for this behavior: 
Above the UCL, the major contributors of shear production terms in the budget equations of $\langle \overline{u_1^\prime u_3^\prime}\rangle_o $ and $k_o$ are 
\begin{equation}
\langle\overline{\mathcal{P}}\rangle_{13,o}= -2\langle \overline{u_3^\prime u_3^\prime}\rangle_l \frac{\partial \langle \overline{u}_1 \rangle_o}{\partial x_3}-2\langle \overline{u_3^\prime u_3^\prime}\rangle_o \frac{\partial \langle \overline{u}_1 \rangle_l }{\partial x_3}
	\label{eq:prod_uw}
\end{equation}
and 
\begin{equation}
\langle\overline{\mathcal{P}}\rangle_{k,o} = -\langle \overline{u_1^\prime u_3^\prime}\rangle_l \frac{\partial \langle \overline{u}_1 \rangle_o}{\partial x_3}-\langle \overline{u_1^\prime u_3^\prime}\rangle_o \frac{\partial 
 \langle\overline{u}_1\rangle_l}{\partial x_3} \ ,
	\label{eq:prod_k}
\end{equation}
respectively.
As the oscillating shear rate travels upwards away from the UCL, it interacts with the local turbulence by modulating $\langle\overline{\mathcal{P}}\rangle_{13,o}$ and $\langle\overline{\mathcal{P}}\rangle_{k,o}$, ultimately yielding the observed oscillations in resolved Reynolds stresses.
On the other hand, no pulsatile-forcing-related terms appear in the budget equations of resolved Reynolds stresses.
This indicates that the oscillating shear rate induced by the pulsatile forcing modifies the turbulence production above the UCL, rather than the pressure forcing itself.
A similar point about pulsatile flows was made in \cite{scotti2001numerical}, where it was stated that ``[...]in the former [pulsatile flow] it is the shear generated at the wall that affects the flow.''
It is worth noting that such a study was, however, based on pulsatile flow over smooth surfaces and at a relatively low Reynolds number. 

In addition, a visual comparison of the contours of $ {\partial \langle \overline{u}_1\rangle_o }/{\partial x_3}$ and $-\langle \overline{u_1^\prime u_3^\prime}\rangle_o $ highlights the presence of a phase lag between such quantities throughout the flow field.
Further examination of this phase lag can be found in \cite{li2023mean}.
During the pulsatile cycle, the turbulence is hence not in equilibrium with the mean flow. This is the case despite the fact that neither the pulsatile forcing nor the induced oscillating shear wave significantly alters the longtime averaged turbulence intensity, as evidenced in figure \ref{fig:longtime_rey}.
To gain further insight into this behavior, the next section examines the structure of turbulence under this non-equilibrium condition. 

\subsection{Quadrant analysis}\label{sec:res-qa}
The discussions will first focus on the impact of flow pulsation on the $u_1^\prime u_3^\prime$ quadrants, with a focus on the ISL. 
This statistical analysis enables the quantification of contributions from different coherent motions to turbulent momentum transport.
The quadrant analysis technique was first introduced by \cite{wallace1972wall}, and has thereafter been routinely employed to characterize the structure of turbulence across a range of flow systems \citep{wallace2016quadrant}. 
The approach maps velocity fluctuations to one of four types of coherent motions (quadrants) in the $u_1^\prime-u_3^\prime$ phase space, namely, 
\begin{equation}
    \begin{cases}
      Q1: & u_1^\prime>0, u_3^\prime>0\ , \\
      Q2: & u_1^\prime<0, u_3^\prime>0\ , \\
      Q3: & u_1^\prime<0, u_3^\prime<0\ , \\
      Q4: & u_1^\prime>0, u_3^\prime<0\ .
    \end{cases}   
    \label{eq:QA_def}
\end{equation}

\begin{figure}
  \centerline{\includegraphics[width=\textwidth]{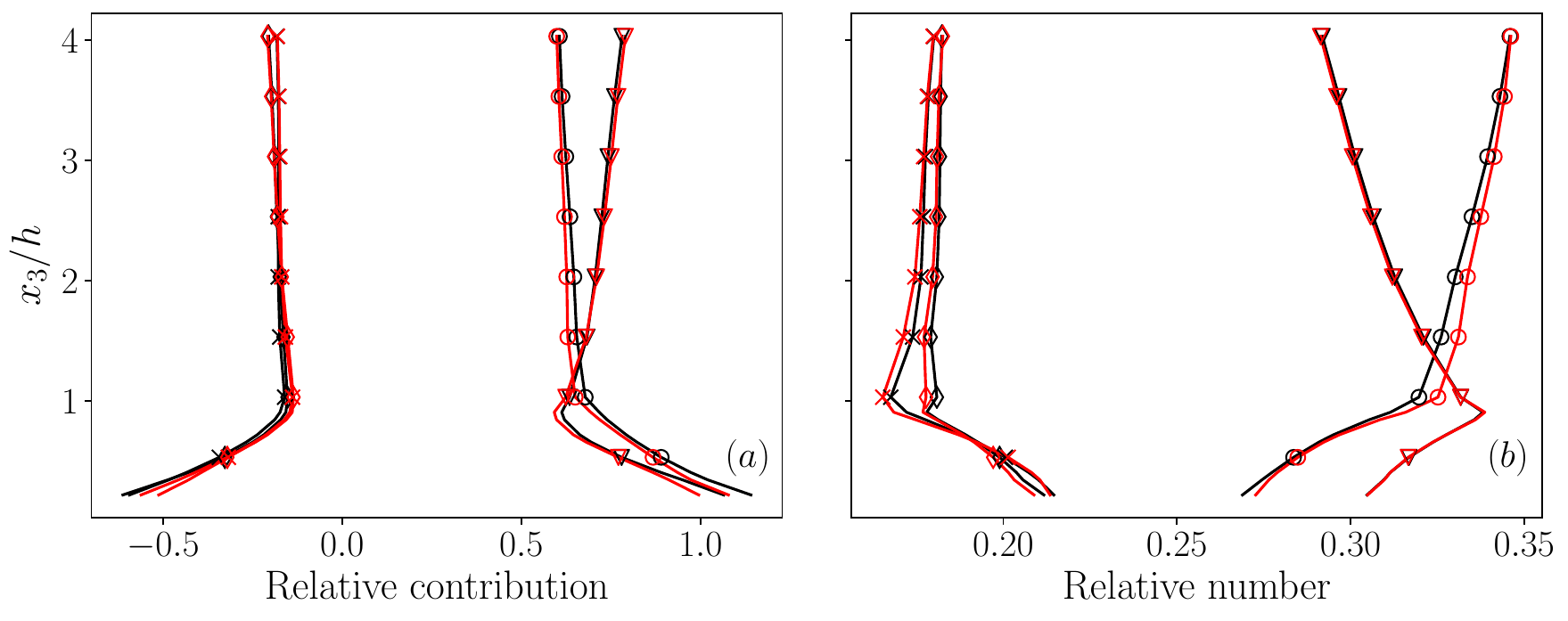}}
  \caption{(\textit{a}) Relative contribution to $\overline{u_1^\prime u_3^\prime}$ by events in each quadrant summed over the wall-parallel planes and the whole sampling time period and (\textit{b}) relative number of events in each quadrant from the PP case (black) and CP (red) as a function of $x_3$. Cross: outward interaction; triangles: ejection; diamonds: inward interaction; circles: sweep.}
\label{fig:QA_all}
\end{figure}

Q2 and Q4 are typically referred to as ejections and sweeps, respectively.
They are the main contributors to the Reynolds shear stress, and constitute the majority of the events in boundary layer flows.
Ejections are associated with the lift-up of low-momentum fluid by vortex induction between the legs of hairpin structures, whereas sweeps correspond to the down-draft of the high-momentum fluid \citep{adrian2000vortex}.
Q1 and Q3 denote outward and inward interactions, and play less important roles in transporting momentum when compared to Q2 and Q4.
\cite{coceal2007structure} and \cite{finnigan2000turbulence} showed that the RSL of stationary flows is dominated by ejections in terms of the number of events, but the overall Reynolds stress contribution from sweep events exceeds that of ejections.
This trend reverses in the ISL.
This behavior is indeed apparent from figure \ref{fig:QA_all}, where ejection and sweep profiles are shown for the CP case (red lines).

We first examine the overall frequency of events in each quadrant and the contribution of each quadrant to the resolved Reynolds shear stress.
For the considered cases, the contribution to $\overline{u_1^\prime u_3^\prime}$ and the number of the events of each quadrant are summed over different wall-parallel planes and over the whole sampling time period (i.e., these are longtime-averaged quantities). 
Results from this operation are also shown in figure \ref{fig:QA_all}.
What emerges from this analysis is that flow pulsation does not significantly alter the relative contribution and frequency of each quadrant. 
Some discrepancies between CP and PP profiles can be observed immediately above the UCL, but do not sum to more than 4\% at any given height.

\begin{figure}
  \centerline{\includegraphics[width=\textwidth]{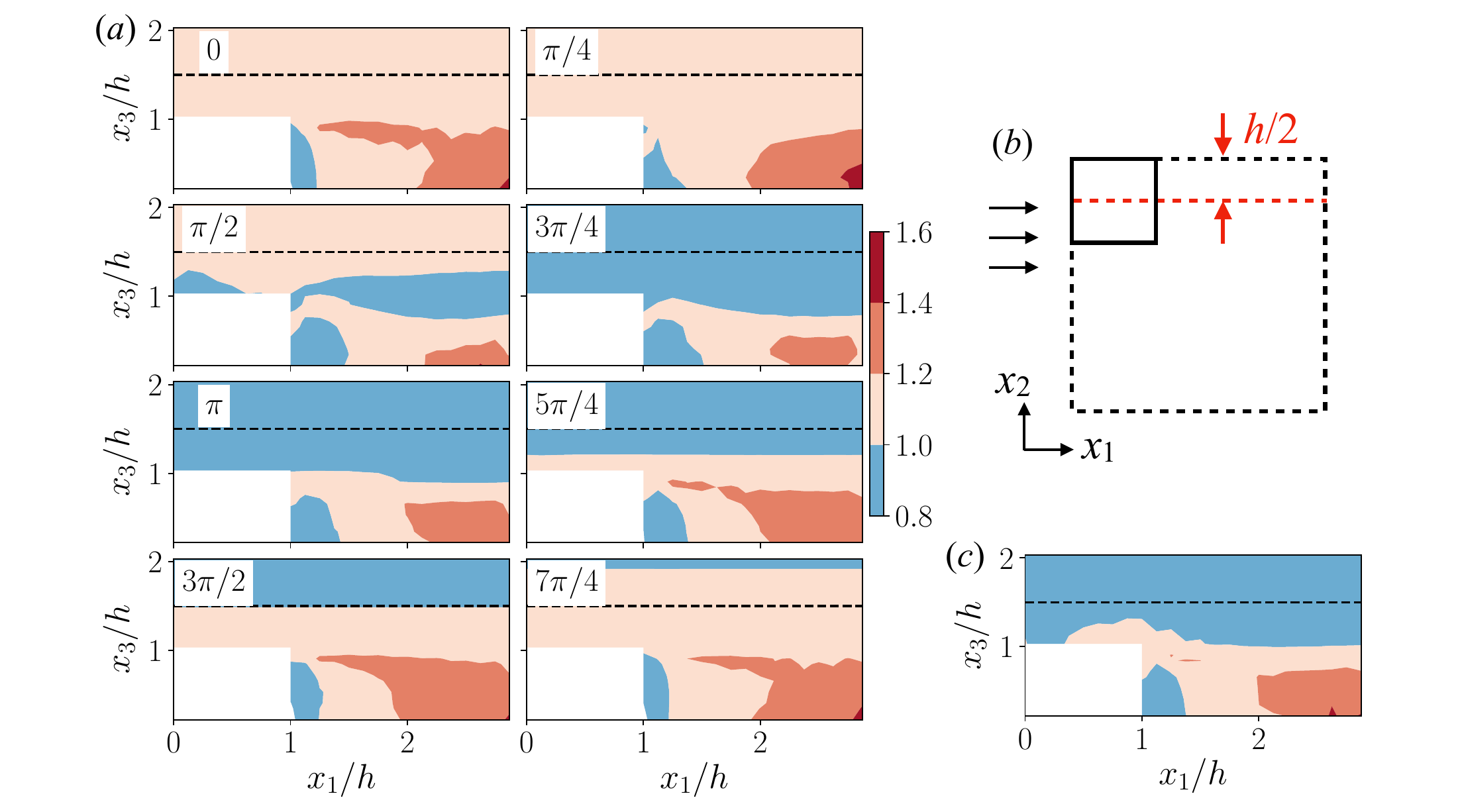}}
 \caption{(\textit{a}) Ratio between the numbers of ejections to sweeps ($\gamma_\#$) from the PP case on a streamwise/wall-normal plane. (\textit{b}) Location of the selected streamwise/wall-normal plane (red dashed line) within a repeating unit. (\textit{c}) $\gamma_\#$ from the CP case on the same plane. Black dashed lines denote $x_3/h=1.5$, where is the upper limit of the RSL.}
\label{fig:QA_ratio_rsl}
\end{figure}

A more interesting picture of the flow field emerges if we consider the phase-dependent behavior of ejections and sweeps.
Hereafter the ratio between the numbers of ejections and sweeps is denoted by $\gamma_\#$, and the ratio of their contribution to $\overline{u_1^\prime u_3^\prime}$ by $\gamma_c$. 
As outlined in the previous section, turbulent fluctuations are defined as deviations from the local ensemble average. 
Consequently, both the frequency of occurrences and the contribution to $\overline{u_1^\prime u_3^\prime}$ from each quadrant are influenced by two main factors: the relative position to the cube within the repeating unit and the phase in the PP case. This dual dependency extends to $\gamma_\#$ and $\gamma_c$ as well.
Conversely, in the CP case, $\gamma_\#$ and $\gamma_c$ are only functions of the spatial location relative to the cube.
Figure \ref{fig:QA_ratio_rsl}(\textit{a},\textit{c}) present $\gamma_\#$ up to $x_3/h=2$ at a selected streamwise/wall-normal plane for the PP and CP cases, respectively. 
The chosen plane cuts through the center of a cube in the repeating unit, as shown in \ref{fig:QA_ratio_rsl}(\textit{b}).
In the cavity, the ejection-sweep pattern from the PP case is found to be qualitatively similar to its CP counterpart throughout the pulsatile cycle (compare subplots (\textit{a},\textit{c}) in figure \ref{fig:QA_ratio_rsl}). 
Specifically, a preponderance of sweeps characterizes a narrow region in the leeward side of the cube (the streamwise extent of this region is $\lessapprox 0.3h$), whereas ejections dominate in the remainder of the cavity. 
As also apparent from figure \ref{fig:QA_ratio_rsl}(\textit{a}), the streamwise extent of the sweep-dominated region features a modest increase (decrease) during the acceleration (deceleration) time period.
During the acceleration phase, the shown above canopy region $(h<x_3<2h)$ transitions from an ejection-dominated flow regime to a sweep-dominated one, and vice versa as the flow decelerates.
This transition initiates just above the cavity, characterized by a higher occurrence of sweeps during the acceleration phase and a predominance of ejections in the deceleration period. 
This continues until both phenomena are distributed throughout the RSL. 
While not discussed in this work, it is worth noting that the trend observed for $\gamma_c$ is precisely the inverse.

\begin{figure}
  \centerline{\includegraphics[width=\textwidth]{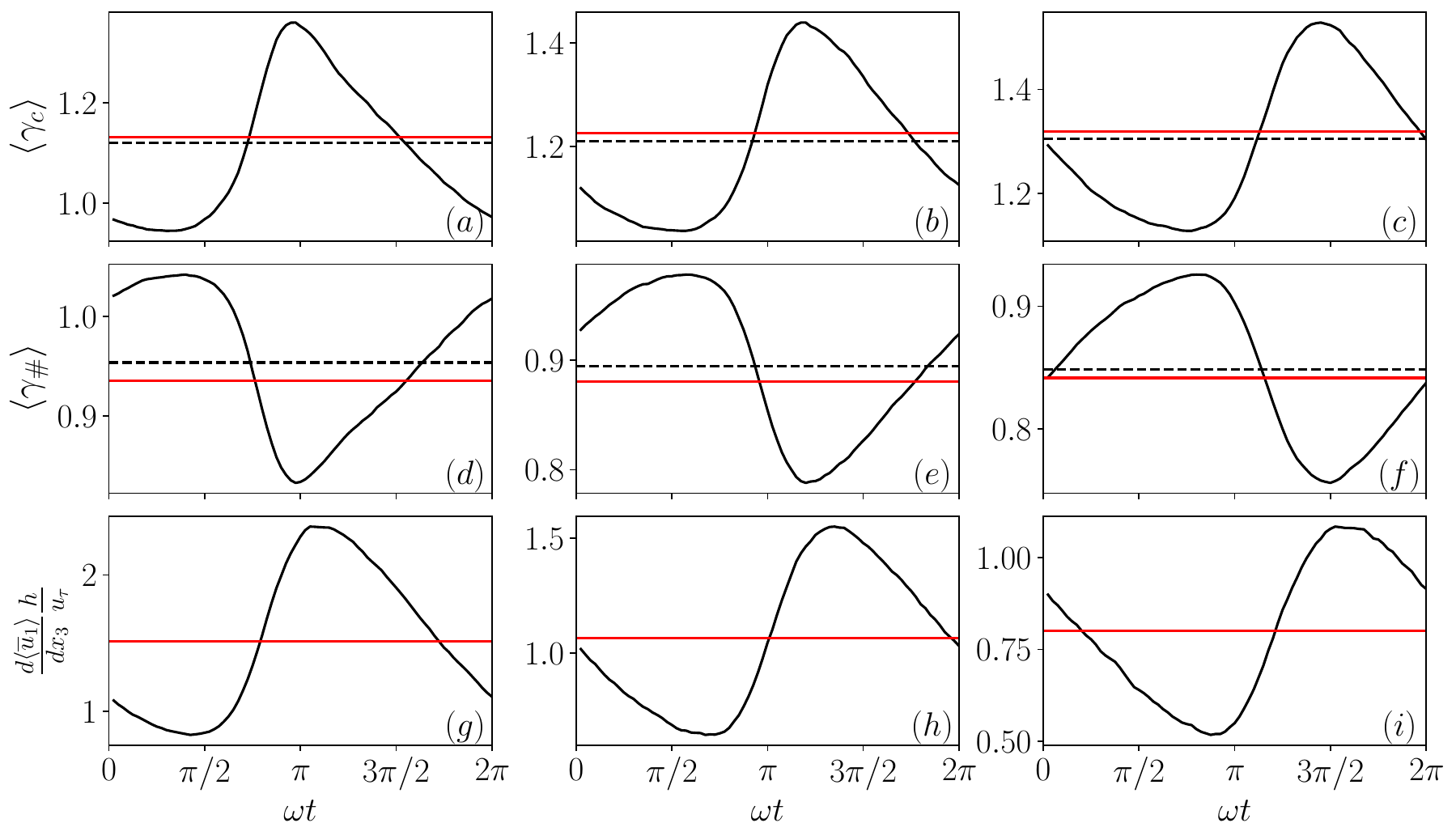}}
\caption{(\textit{a}) - (\textit{c}): Intrinsic-averaged ratio of contributions to $\overline{ u_1^\prime u_3^\prime }$ from ejections and sweeps ($\langle \gamma_c \rangle$); (\textit{d}) - (\textit{f}): intrinsic-averaged ratio of ejections to sweeps ($\langle \gamma_\# \rangle$); (\textit{g}) - (\textit{i}): intrinsic and phase-averaged shear rate $ {\partial \langle \overline{u}_1 \rangle}/{\partial x_3}$ from the PP case at three wall-normal locations within the ISL (\textit{a},\textit{d},\textit{g}) $x_3/h=2$, (\textit{b},\textit{e},\textit{h}) $x_3/h=3$ and (\textit{c},\textit{f},\textit{i}) $x_3/h=4$ as a function of phase. Black dashed lines denote longtime-averaged values, whereas solid red lines represent corresponding quantities from the CP case.}
\label{fig:QA_ratio_isl}
\end{figure}

\begin{figure}
  \centerline{\includegraphics[width=\textwidth]{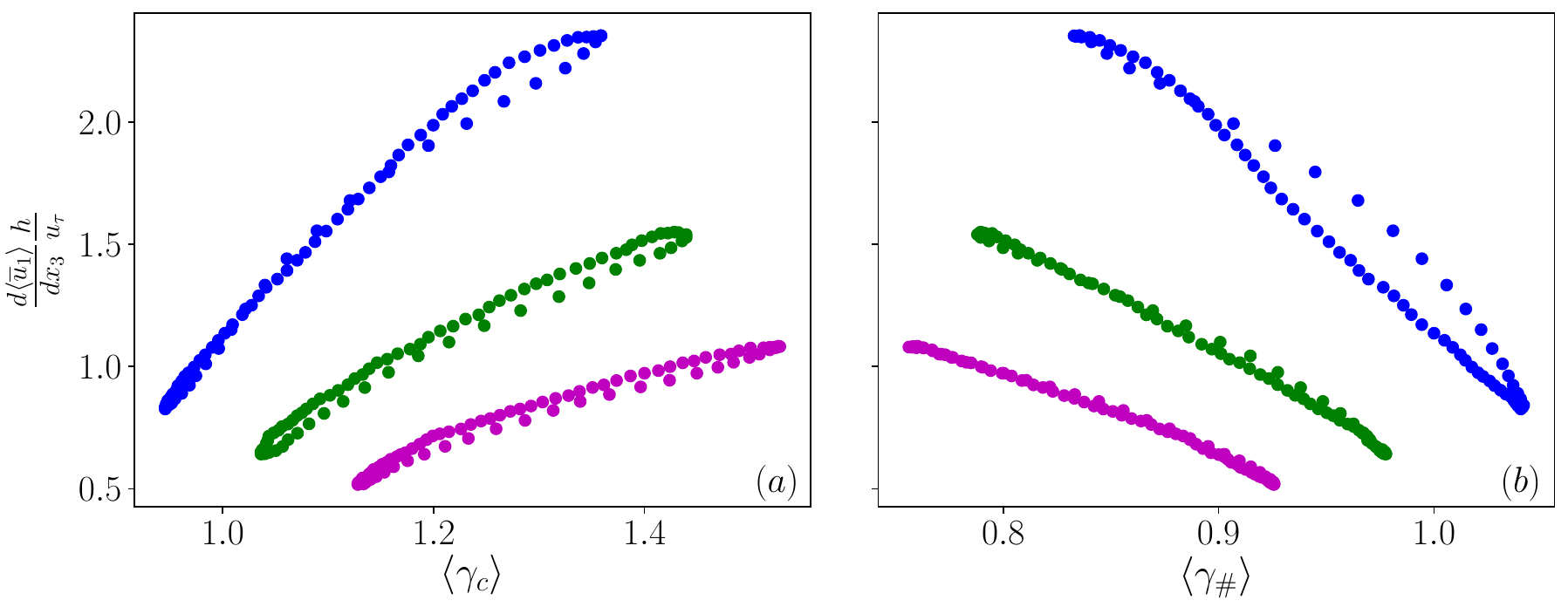}}
\caption{(\textit{a}) $\langle \gamma_c \rangle$ and (\textit{b}) $\langle \gamma_\# \rangle$ versus $ {\partial \langle \overline{u}_1 \rangle}/{\partial x_3}$ at $x_3/h=2$ (blue), $x_3/h=3$ (green) and $x_3/h=4$ (magenta).}
\label{fig:QA_vs_dudz}
\end{figure}

Shifting the attention to the ejection-sweep pattern in the ISL, which is indeed the main focus of this study, figure \ref{fig:QA_ratio_isl} shows the intrinsic average of $\gamma_c$ and $\gamma_\#$ in the $x_3/h = \{2, 3, 4 \}$ planes.
These quantities are hereafter denoted as $\langle \gamma_c \rangle$ and $\langle \gamma_\# \rangle$, respectively.
The use of $\langle \gamma_c \rangle$ and $\langle \gamma_\# \rangle$ instead of $\gamma_c$ and $\gamma_\#$ to characterize the ejection-sweep pattern in the ISL can be justified by the fact that the spatial variations in $\gamma_\#$ and $\gamma_c$ on the wall-parallel directions vanish rapidly above the RSL, as apparent from figure \ref{fig:QA_ratio_rsl}. 
This is in line with the observations of \cite{kanda2004large} and \cite{castro2006turbulence} that the spatial variations in $\gamma_\#$ and $\gamma_c$ are concentrated in the RSL for stationary flow over urban canopy. 
Further, as shown in figure \ref{fig:QA_ratio_isl}, the ejection-sweep pattern varies substantially during the pulsatile cycle.
For instance, at a relative height of $x_3/h=2$, even though the contribution from ejections to $\overline{ u_1^\prime u_3^\prime }$ dominates in a longtime average sense ($\langle \gamma_c \rangle_l> 1$), sweeps contributions prevail  for $\omega t \in [0, \pi/2]$. 
Interestingly, at a given wall-normal location, this ejection-sweep pattern appears to be directly controlled by the intrinsic and phase-averaged shear rate $ {\partial \langle \overline{u}_1 \rangle}/{\partial x_3}$.
This is particularly evident when $\langle \gamma_c \rangle$ and $\langle \gamma_\# \rangle$ are plotted against $ {\partial \langle \overline{u}_1 \rangle}/{\partial x_3}$ (refer to figure \ref{fig:QA_vs_dudz}). 
As $ {\partial \langle \overline{u}_1 \rangle}/{\partial x_3}$ increases at a given $x_3$, the corresponding $\langle \gamma_c \rangle$ increases whereas $\langle \gamma_\# \rangle$ decreases, highlighting the presence of fewer but stronger ejections events. 
Maxima and minima of $\langle \gamma_c \rangle$ and $\langle \gamma_\# \rangle$ approximately coincide with the maxima of ${\partial \langle \overline{u}_1 \rangle}/{\partial x_3}$.
This observation is consistent across the considered planes. 
As discussed in the next sections, such behavior can be attributed to time variations in the geometry of ISL structures.

\subsection{Spatial and temporal flow coherence}\label{sec:res-Rii}

To gain a better understanding of the extent and organization of coherent structures in the ISL, this section analyzes two-point velocity autocorrelation maps. 
These flow statistics provide information on the correlation of the flow field in space, making it an effective tool for describing spatial flow coherence \citep{dennis2011experimental, guala2012vortex}.
For the PP case, the phase-dependent two-point correlation coefficient tensor $\overline{R}_{ij}$ can be defined as
\begin{equation}
  \overline{R}_{ij}(\Delta_1,\Delta_2,x_3,x_3^*,t)=\frac{\langle \overline{ u_i^\prime(x_1,x_2,x_3^*,t) u_j^\prime(x_1+\Delta_1,x_2+\Delta_2,x_3,t)} \rangle}{\sqrt{\langle \overline{u_i^\prime u_i^\prime}\rangle (x_3^*,t)\langle \overline{ u_j^\prime u_j^\prime}\rangle (x_3,t)}} \ ,
    \label{eq:rij}
\end{equation}
where $\Delta_i$ is the separation on the wall-parallel directions, $x_3^*$ represents a reference wall-normal location, and $t$ denotes the phase.
In the CP case, the flow is statistically stationary, and therefore $\overline{R}_{ij}$ is not a function of $t$, i.e., $\overline{R}_{ij} = \overline{R}_{ij,l}$.

\begin{figure}
  \centerline{\includegraphics[width=\textwidth]{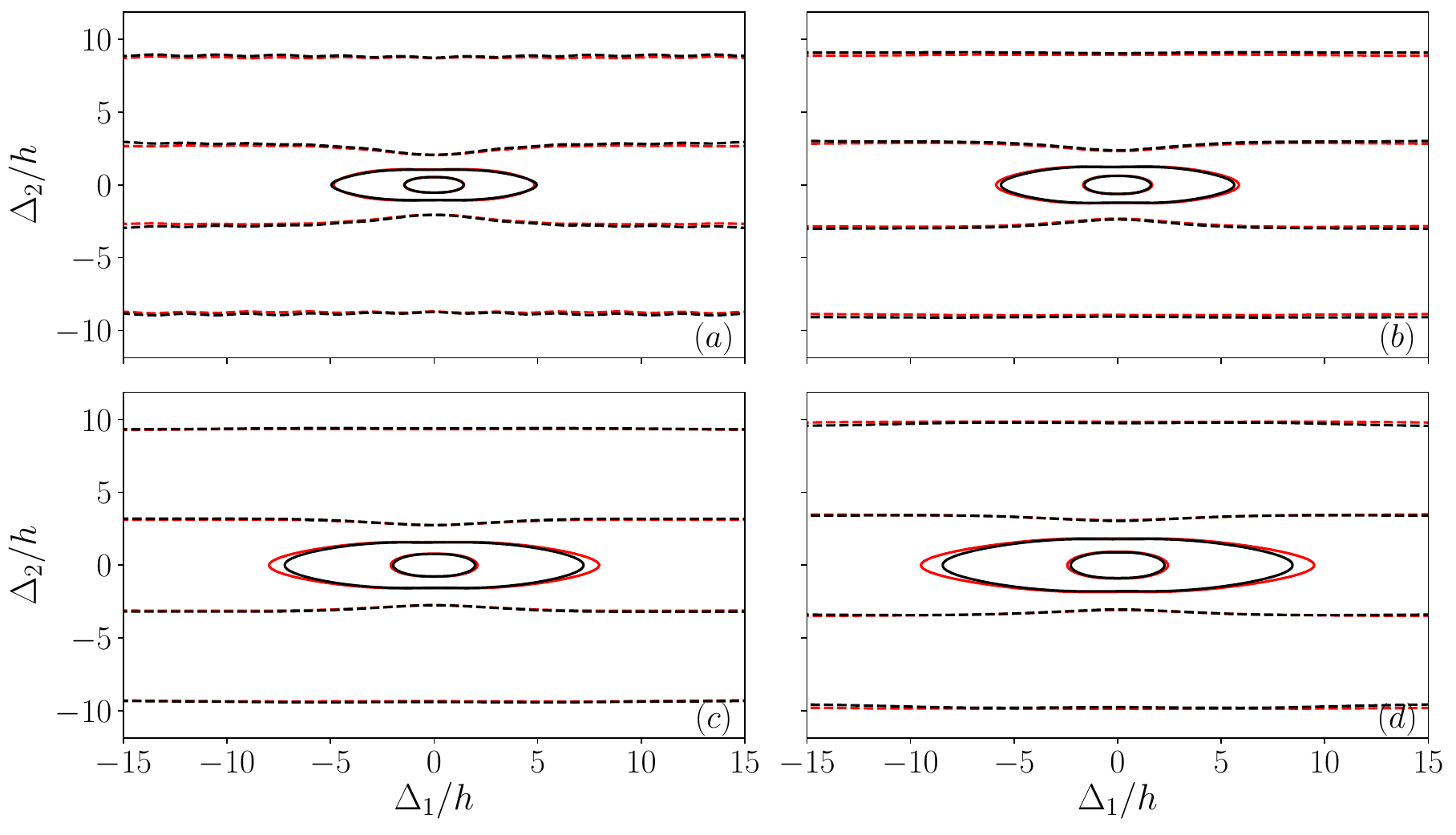}}
  \caption{Longtime-averaged two-point correlation coefficient tensor $\overline{R}_{11,l}$ at (\textit{a}) $x_3^*/h=1.5$, (\textit{b}) $x_3^*/h=2$, (\textit{c}) $x_3^*/h=3$, and (\textit{d}) $x_3^*/h=4$. Black lines correspond to the PP case, and red lines to the CP one. $\overline{R}_{11,l}=0.6$ and $\overline{R}_{11,l}=0.3$ are denoted by solid lines, and dashed lines represent $\overline{R}_{11,l}=0$.}
\label{fig:ruu_bar_xy}
\end{figure}

\begin{figure}
  \centerline{\includegraphics[width=\textwidth]{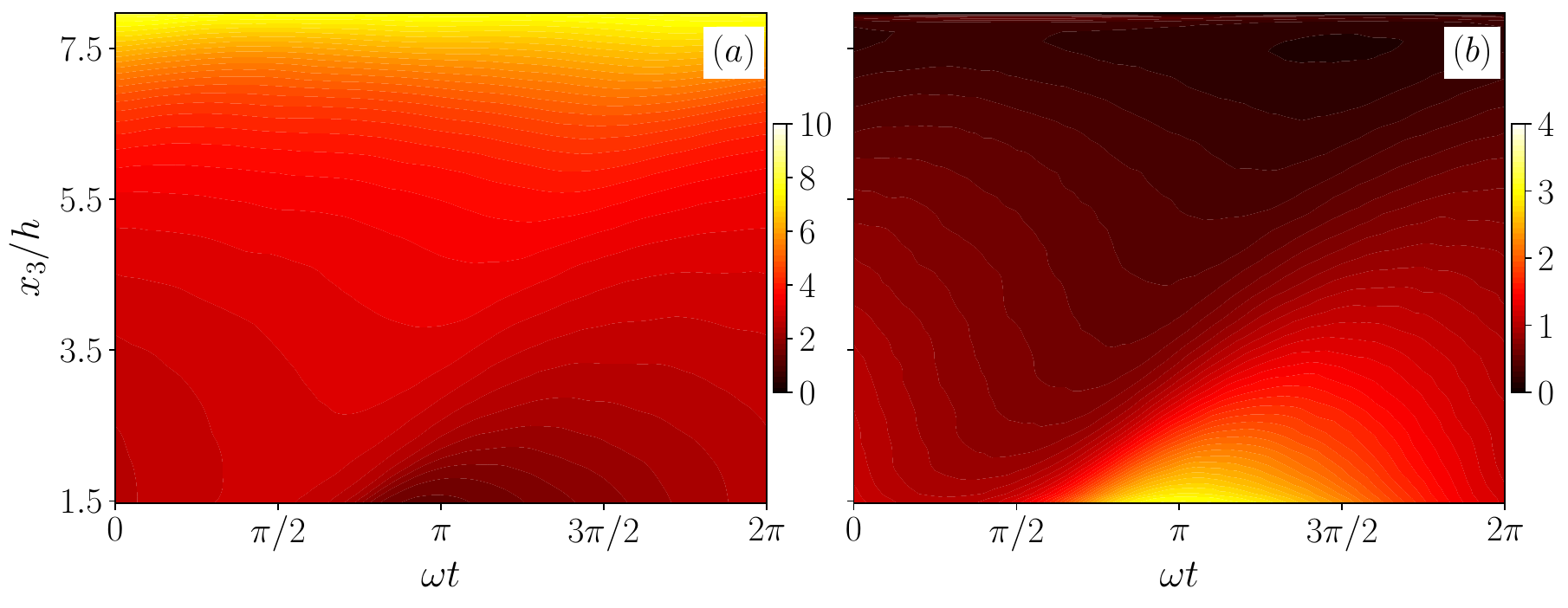}}
  \caption{Time evolution of (\textit{a}) the cross-stream streak width normalized by $h$ and (\textit{b}) $ \partial \langle \overline{u}_1 \rangle/\partial x_3$. The cross-stream width is identified as the first zero crossing of the $\overline{R}_{11}=0$ field.}
\label{fig:ruu_xy_y0}
\end{figure}

Figure \ref{fig:ruu_bar_xy} compares $\overline{R}_{11,l}$ for the PP and CP cases over the $x_3^*/h = \{1.5, 2, 3, 4 \}$ planes. 
In both cases, $\overline{R}_{11,l}$ features an alternating sign in the cross-stream direction, signaling the presence of low- and high-momentum streaks flanking each other in the cross-stream direction.
The cross-stream extent of longtime-averaged streaks can be identified as the first zero-crossing of the $\overline{R}_{11,l}$ contour in the $\Delta_2$ direction. 
Based on this definition, figure \ref{fig:ruu_bar_xy} shows that flow unsteadiness has a modest impact on such a quantity. 
This finding agrees with observations from \cite{zhang2019experimental} for pulsatile flow over smooth surfaces.
Further, although not shown, the streamwise and cross-stream extent of streaks increases linearly in $x_3$, suggesting that Townsend's attached-eddy hypothesis is valid in a longtime average sense \citep{marusic2019attached}.

Turning the attention to the phase-averaged flow field, figure \ref{fig:ruu_xy_y0} shows the time variation of the cross-stream streaks extent, which is identified as the first zero crossing of the $\overline{R}_{11}=0$ field in the cross-stream direction. 
The linear $x_3$-scaling of the streak width breaks down in a phase-averaged sense.
Such a quantity indeed varies substantially during the pulsatile cycle, diminishing in magnitude as $ {\partial \langle \overline{u}_1 \rangle}/{\partial x_3}$ increases throughout the boundary layer. 
Interestingly, when ${\partial \langle \overline{u}_1 \rangle}/{\partial x_3}$ reaches its maximum at $\omega t \approx \pi$ and $x_3/h \approx 1.5$, the cross-stream extent of streaks approaches zero, suggesting that streaks may not be a persistent feature of pulsatile boundary layer flows.

\begin{figure}
  \centerline{\includegraphics[width=\textwidth]{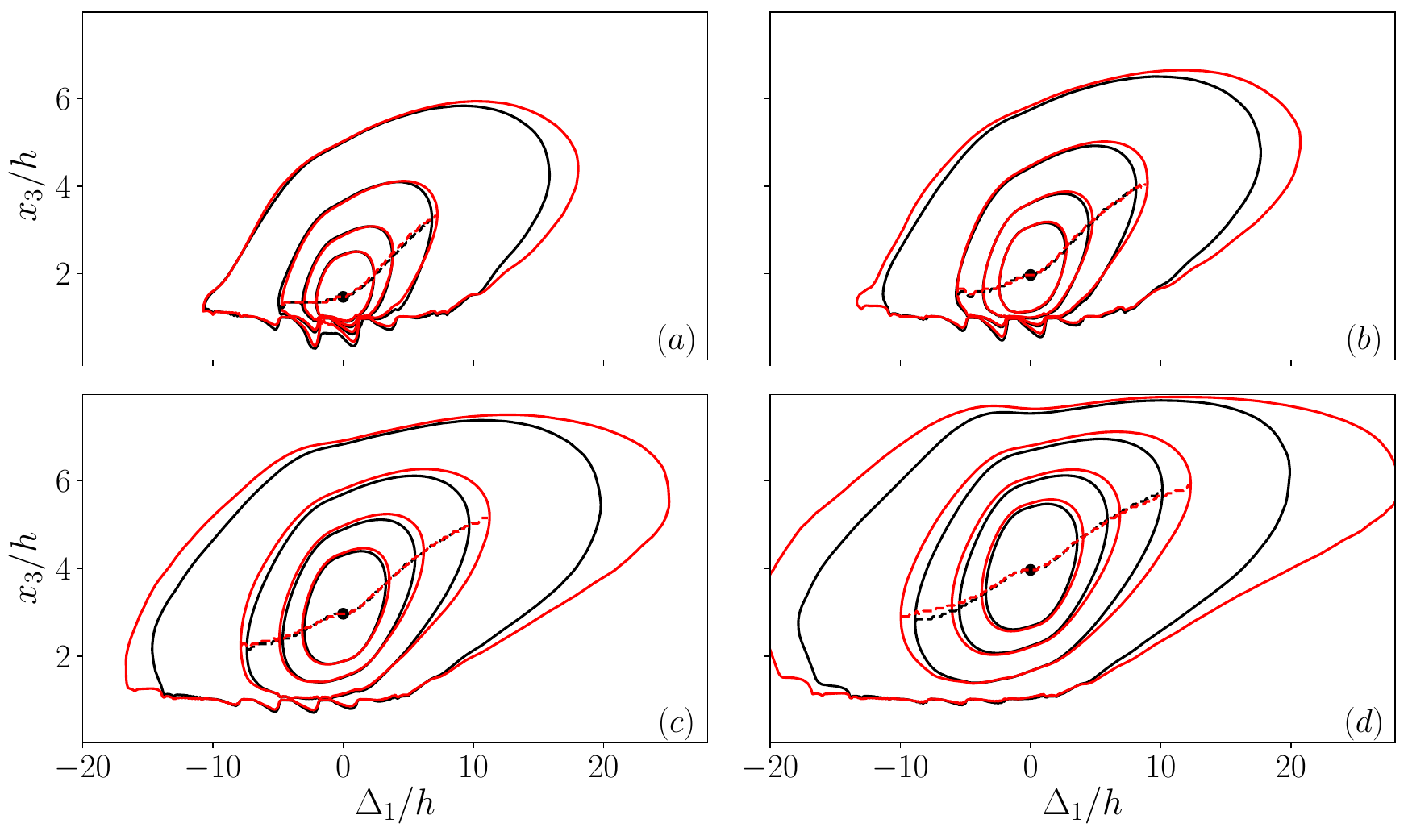}}
  \caption{$\overline{R}_{11,l}$ in the streamwise/wall-normal plane of the PP (black) and CP (red) cases. Results correspond to four reference wall-normal locations: (\textit{a}) $x_{3}^*/h=1.5$, (\textit{b}) $x_{3}^*/h=2$, (\textit{c}) $x_{3}^*/h=3$, and (\textit{d}) $x_{3}^*/h=4$. Contour levels (solid lines) range from $0.2$ to $0.5$ with increments of $0.1$.
  Dashed lines denote the locus of the maximum correlation at each streamwise location. The slopes of the dashed lines represent the tilting angles of the structures.}
\label{fig:ruu_bar_xz}
\end{figure}

To further quantify topological changes induced by flow pulsation, we hereafter examine variations in the streamwise and wall-normal extent of coherent structures.
Such quantities will be identified via the $\overline{R}_{11}=0.3$ contour, in line with the approach used by \cite{krogstad1994structure}.
Note that the choice of the $\overline{R}_{11}$ threshold for such a task is somewhat subjective, and several different values have been used in previous studies to achieve this same objective, including $\overline{R}_{11}=0.4$ \citep{takimoto2013length} and $\overline{R}_{11}=0.5$ \citep{volino2007turbulence, guala2012vortex}. 
In this study, the exact threshold is inconsequential as it does not impact the conclusions.
Figure \ref{fig:ruu_bar_xz} presents $\overline{R}_{11,l}$ contours in the streamwise/wall-normal plane for $x_3^*/h = \{ 1.5,2,3,4 \}$.
The jagged lines at $x_3/h \approx 1$ (the top of the UCL) bear the signature of roughness elements.
The dashed lines passing through $x_3^*$ identify the locus of the maxima in $\overline{R}_{11,l}$ at each streamwise location. 
The inclination angle of such lines can be used as a surrogate for the longtime-averaged tilting angle of the coherent structure \citep{chauhan2013structure,salesky2020revisiting}. 
It is clearly observed that at each reference wall-normal location, the tilting angle of longtime-averaged structures is similar between the PP case and CP.
The tilting angle in both cases decreases monotonically and slowly from $15^\circ$ at $x_3^*/h=1.5$ to $10^\circ$ at $x_3^*/h=4$---a behavior that is in excellent agreement with results from \cite{coceal2007structure}, even though a different urban canopy layout was used therein. 
Further, the identified tilting angle is also similar to the one inferred from real-world ABL observations in \cite{hutchins2012towards} and \cite{chauhan2013structure}.
On the other hand, longtime-averaged coherent structures in the PP case are relatively smaller than in the CP case in both the streamwise and wall-normal coordinate directions.
Discrepancies become more apparent with increasing $x_3^*$. 
Specifically, the difference in the streamwise extent of the longtime-averaged structure from the two cases increases from $2\%$ at $x_3^*/h=1.5$ to $15\%$ at $x_3^*/h=4$.
Corresponding variations in the wall-normal extent are $2\%$ and $4\%$.  

\begin{figure}
  \centerline{\includegraphics[width=\textwidth]{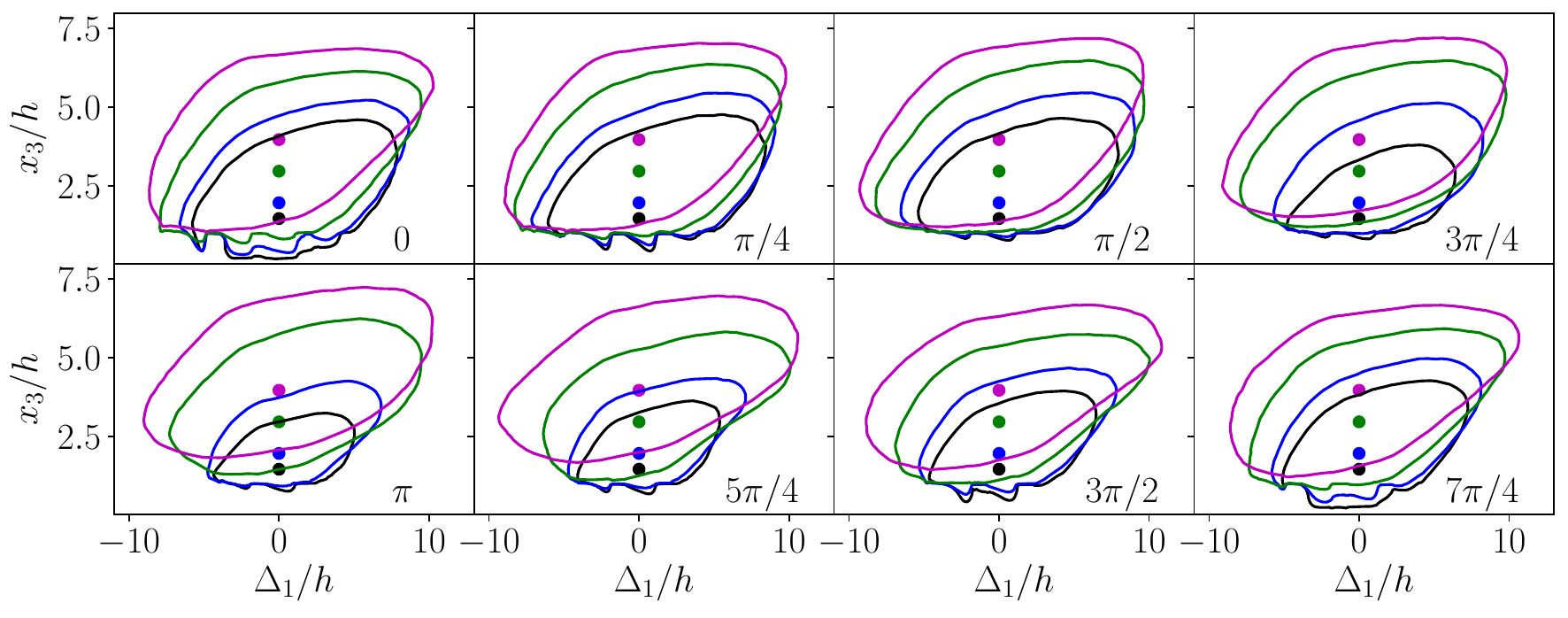}}
  \caption{Time evolution of $\overline{R}_{11}=0.3$ in the streamwise/wall-normal plane. Line colors denote the contours corresponding to different $x_3^*$ planes: $x_3^*/h=1.5$ (black), $x_3^*/h=2$ (blue), $x_3^*/h=3$ (green), and $x_3^*/h=4$ (magenta). Dots highlight the location of the reference plane.}
\label{fig:ruu_xz_phase}
\end{figure}

\begin{figure}
  \centerline{\includegraphics[width=0.5\textwidth]{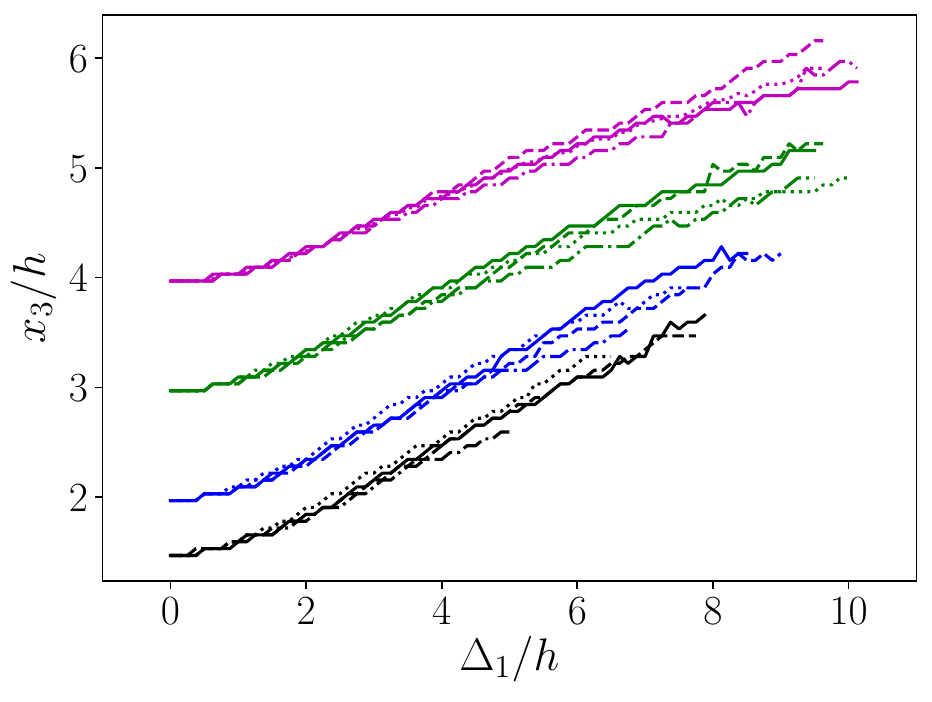}}
  \caption{The locus of the maximum $\overline{R}_{11}$ at four phases: $\omega t=0$ (solid lines), $\omega t=\pi/2$ (dashed lines), $\omega t=\pi$ (dashed dotted lines), and $\omega t=3\pi/2$ (dotted lines). Line colors denote different reference elevations: $x_3^*/h=1.5$ (black), $x_3^*/h=2$ (blue), $x_3^*/h=3$ (green), and $x_3^*/h=4$ (magenta).}
\label{fig:ruu_xz_phase_angle}
\end{figure}

More insight into the mechanisms underpinning the observed behavior can be gained by examining the time evolution of such structures for the PP case in figure \ref{fig:ruu_xz_phase}.
When taken together with figure \ref{fig:ruu_xy_y0}(\textit{b}), it becomes clear that both the streamwise and the wall-normal extents of the coherent structures tend to reduce with increasing local $ \partial \langle \overline{u}_1 \rangle/\partial x_3$. 
Compared to the streamwise extent, the wall-normal extent of the coherent structure is more sensitive to changes in $\partial \langle \overline{u}_1 \rangle/\partial x_3$.
For example, at $x_3^*/h=4$, we observe an overall $15\%$ variation in the wall-normal extent of the coherent structure during a pulsation cycle, whereas the corresponding variation in streamwise extent is $8\%$. 
Further, the flow field at the considered heights appears to be more correlated with the flow in the UCL for small $\partial \langle \overline{u}_1 \rangle/\partial x_3$, thus highlighting a stronger coupling between flow regions in the wall-normal direction.
Interestingly, the tilting angle of the coherent structure remains constant during the pulsatile cycle, as shown in figure \ref{fig:ruu_xz_phase_angle}. 

Next, we will show that the hairpin vortex packet paradigm \citep{adrian2007hairpin} can be used to provide an interpretation for these findings. 
Note that alternative paradigms, such as that proposed by \cite{del2006self}, may offer different interpretations of the results, but are not discussed in this work.
The validity of such a paradigm is supported by a vast body of evidence from laboratory experiments of canonical TBL \citep{adrian2000vortex, christensen2001statistical, dennis2011experimental} to ABL field measurements \citep{hommema2003packet, morris2007near} and numerical simulations \citep{lee2011direct,eitel2015hairpin}.
This formulation assumes that the dominant ISL structures are hairpin vortex packets, consisting of a sequence of hairpin vortices organized in a quasi-streamwise direction with a characteristic inclination angle relative to the wall. 
These structures encapsulate the low-momentum regions, also known as ``streaks.''
The structural information obtained from the two-point correlation has been considered to reflect the averaged morphology of the hairpin vortex packets \citep{zhou1999mechanisms,ganapathisubramani2005investigation,volino2007turbulence,hutchins2012towards,guala2012vortex}. 
Specifically, in this study, the observed changes in $\overline{R}_{11,l}$ between the CP and PP cases and of $\overline{R}_{11}$ contours during the pulsatile cycle reflect corresponding changes in the geometry of vortex packets in a longtime- and phase-averaged sense. 
That is, as $ \partial \langle \overline{u}_1 \rangle/\partial x_3$ increases, the phase-averaged size of vortex packets is expected to shrink, and, in the longtime-averaged sense, the vortex packets are smaller than their counterparts in the CP case.
However, upon inspection of $\overline{R}_{11}$ in figure \ref{fig:ruu_xz_phase}, it is unclear whether the observed change in packet size is attributable to variations in the composing hairpin vortices or the tendency for packets to break into smaller ones under high $ \partial \langle \overline{u}_1 \rangle/\partial x_3$ and merge into larger ones under low $ \partial \langle \overline{u}_1 \rangle/\partial x_3$.
To answer this question, we will next examine the instantaneous turbulence structures and extract characteristic hairpin vortices through conditional averaging.
Also, the constant tilting angle of the structure evidenced in figure \ref{fig:ruu_xz_phase_angle} during the pulsatile cycle indicates that, no matter how vortex packets break and reorganize and how individual hairpin vortices deform in response to the time-varying shear rate, the hairpin vortices within the same packet remain aligned with a constant tilting angle. 

\subsection{Instantaneous flow structure}\label{sec:res-inst}

\begin{figure}
  \centerline{\includegraphics[width=\textwidth]{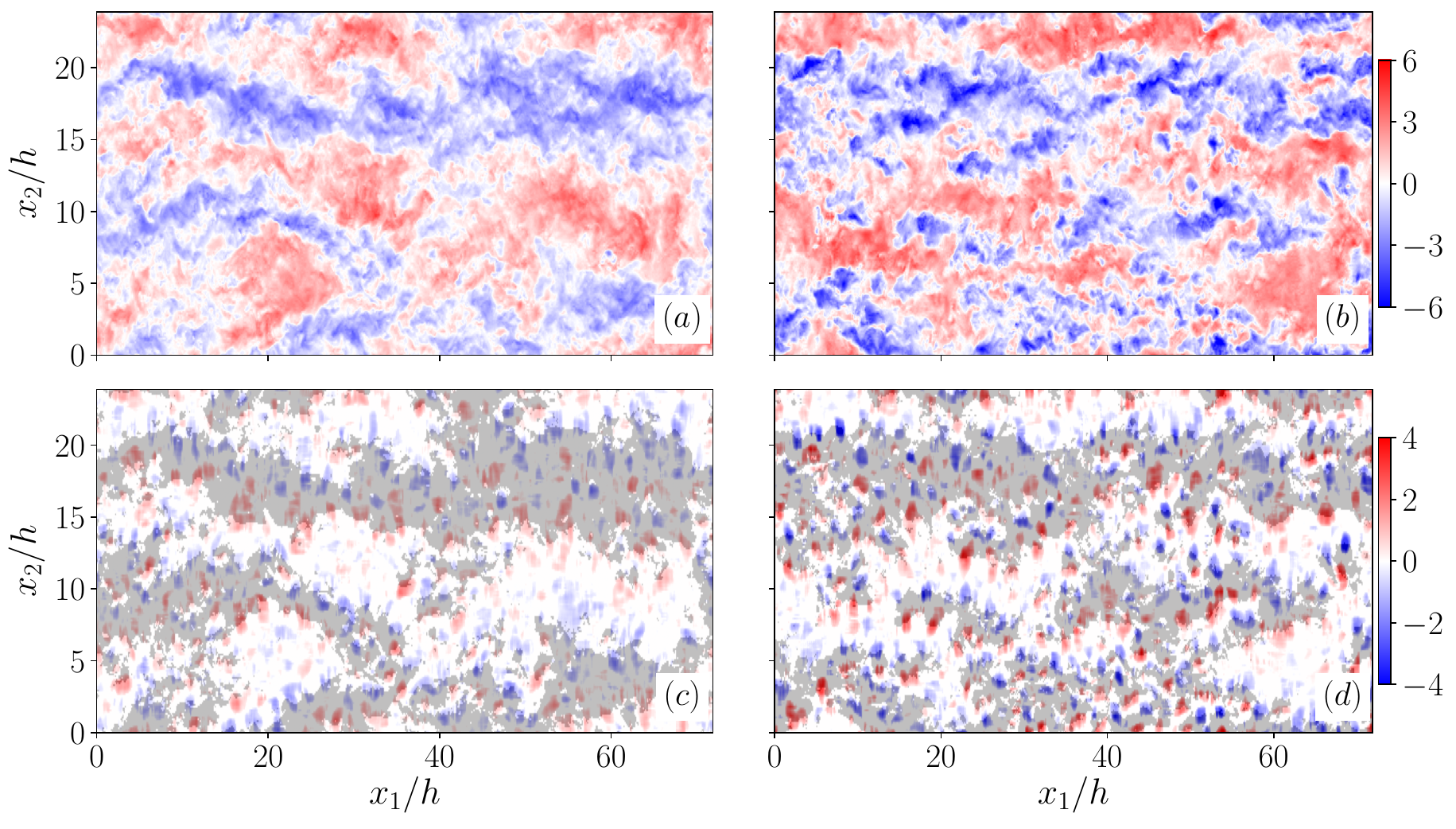}}
  \caption{(\textit{a},\textit{b}): Instantaneous fluctuating streamwise velocity $u_1^\prime$ normalized by ${u}_{\tau}$ at $x_3=2h$; (\textit{c},\textit{d}): wall-normal swirl strength $\lambda_{s,3}$ of the PP case at $x_3=2h$. (\textit{a},\textit{c}): $\omega t=\pi/2$, ; (\textit{b},\textit{d}), $\omega t=\pi$. Shaded regions in (\textit{c},\textit{d}) highlight the low-momentum ($u_1^\prime<0$) regions. The instantaneous flow fields correspond to the same pulsatile cycle. Green solid lines highlight the background location of the cuboids.}
\label{fig:inst_xy_z2p0}
\end{figure}

\begin{figure}
  \centerline{\includegraphics[width=\textwidth]{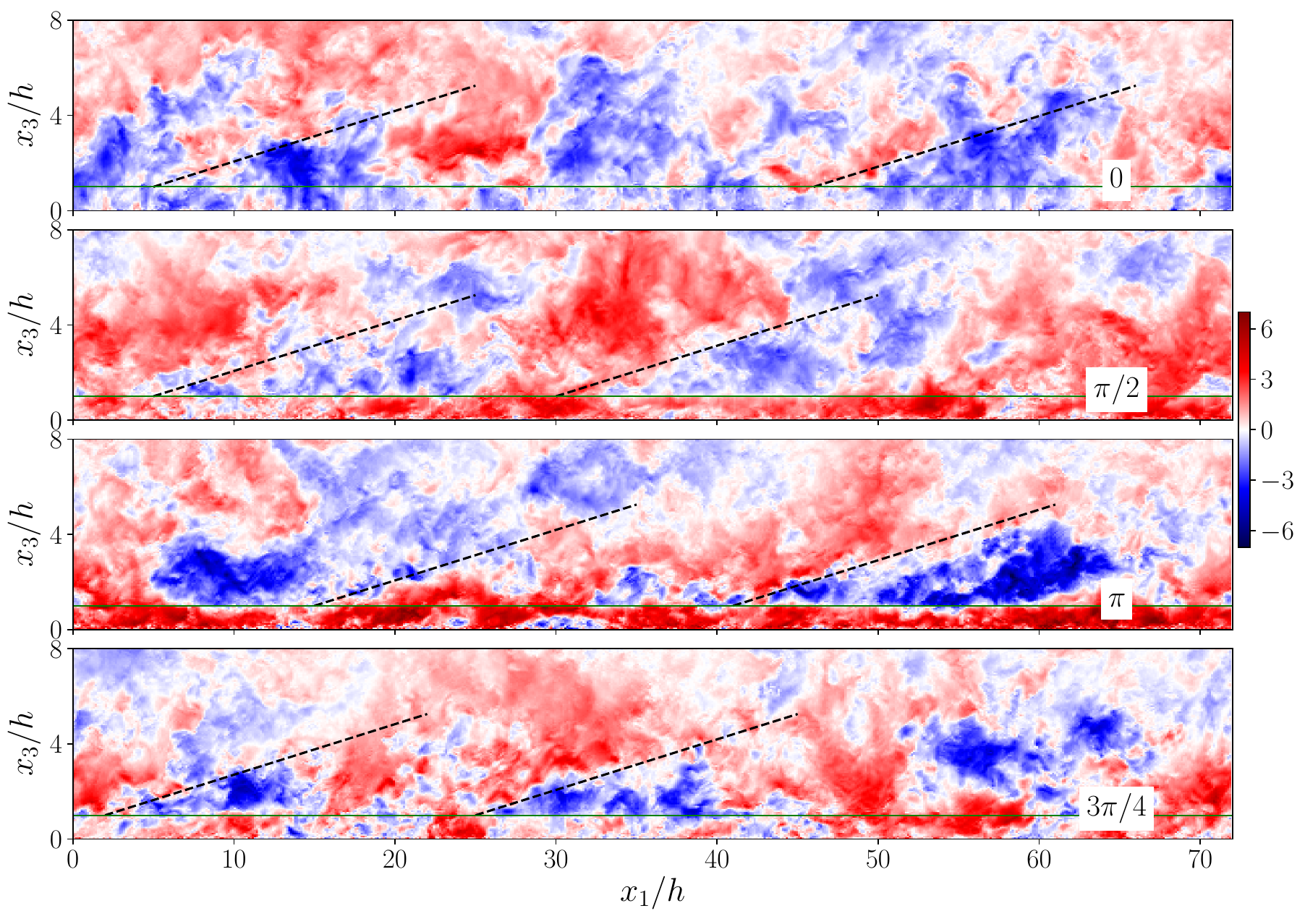}}
  \caption{Instantaneous fluctuating streamwise velocity $u_1^\prime$ in a streamwise/wall-normal plane during a pulsatile cycle. Black dashed lines denote the $12^\circ$ structural tilting angle of the coherent structure. Green solid lines represent the canopy layer top.}
\label{fig:inst_xz}
\end{figure}

Figure \ref{fig:inst_xy_z2p0}(\textit{a},\textit{b}) show the instantaneous fluctuating streamwise velocity $u_1^\prime$ at $x_3/h=1.5$ from the PP case.
The chosen phases, $\omega t=\pi/2$ and $\omega t=\pi$, correspond to the local minimum and maximum of $ \partial \langle \overline{u}_1 \rangle/\partial x_3$, respectively (see figure \ref{fig:QA_ratio_isl},\textit{g}).
Streak patterns can be observed during both phases.
As shown in figure \ref{fig:inst_xy_z2p0}(\textit{a}), at low $ \partial \langle \overline{u}_1 \rangle/\partial x_3$ values, instantaneous $u_1^\prime$ structures intertwine with neighboring ones, and form large streaks with a cross-stream extent of about $5h$. 
Conversely, when $ \partial \langle \overline{u}_1 \rangle/\partial x_3$ is large, the streaks are shrunk into smaller structures, which have a cross-stream extent of about $h$. 
This behavior is consistent with the observations we made based on figure \ref{fig:ruu_xy_y0}.

Further insight into the instantaneous flow field can be gained by considering the low-pass filtered wall-normal swirl strength $\lambda_{s,3}$, shown in figures \ref{fig:inst_xy_z2p0}(\textit{c},\textit{d}).
The definition of the signed planar swirl strength $\lambda_{s,i}$ is based on the studies of \cite{stanislas2008vortical} and \cite{elsinga2012tracking}.
The magnitude of $\lambda_{s,i}$ is the absolute value of the imaginary part of the eigenvalue of the reduced velocity gradient tensor $J_{jk}$, which is 
\begin{equation}
    J_{jk}=\begin{bmatrix}
            {\partial u_{j}}/{\partial x_{j}} & {\partial u_{j}}/{\partial x_{k}}\\
            {\partial u_{k}}/{\partial x_{j}} & {\partial u_{k}}/{\partial x_{k}}
        \end{bmatrix}, i \neq j \neq k \ ,
    \label{eq:mag_lambda}
\end{equation}
with no summation over repeated indices. 
The sign of $\lambda_{s,i}$ is determined by the vorticity component $\omega_i$.
Positive and negative $\lambda_{s,i}$ highlight regions with counterclockwise and clockwise swirling motions, respectively.
To eliminate the noise from the small-scale vortices, we have adopted the \cite{tomkins2003spanwise} idea and low-pass filtered the $\lambda_{s,i}$ field (a compact top-hat filter) with support $h$ to better identify instantaneous hairpin features. 
As apparent from this figure, low-momentum regions are bordered by pairs of oppositely signed $\lambda_{s,3}$ regions at both the considered phases; these counter-rotating rolls are a signature of hairpin legs. 
Based on these signatures, it is also apparent that hairpin vortices tend to align in the streamwise direction.
Comparing subplots (\textit{c},\textit{d}) in figure \ref{fig:inst_xy_z2p0}, it is clear that, as $ \partial \langle \overline{u}_1 \rangle/\partial x_3$ increases, the swirling strength of the hairpin's legs is intensified, which in turn increases the momentum deficits in the low-momentum regions between the hairpin legs. 
This behavior leads to a narrowing of low-momentum regions to satisfy continuity constraints. 
Also, it is apparent that a larger number of hairpin structures populates the flow field at a higher $ \partial \langle \overline{u}_1 \rangle/\partial x_3$, which can be attributed to hairpin vortices spawning offsprings in both the upstream and downstream directions as they intensify \citep{zhou1999mechanisms}. 

Figure \ref{fig:inst_xz} displays a $u_1^\prime$ contour for the PP case at a streamwise/wall-normal plane. 
Black dashed lines feature a tilting angle $\theta=12^\circ$. 
It is evident that the interfaces of the low- and high-momentum regions, which are representative instantaneous manifestations of hairpin packets \citep{hutchins2012towards}, feature a constant tilting angle during the pulsatile cycle.
This behavior is in agreement with findings from the earlier $\overline{R}_{11}$ analysis, which identified the typical tilting angle of coherent structures as lying between $10^\circ$ to $15^\circ$, depending on the reference wall-normal location.
We close this section by noting that while the instantaneous flow field provides solid qualitative insight into the structure of turbulence for the considered flow field, a more statistically representative picture can be gained by conditionally averaging the flow field on selected instantaneous events. 
This will be the focus of the next section.  

\subsection{Temporal variability of the composite hairpin vortex}\label{sec:res-condavg}

This section aims at providing more quantitative insights into the temporal variability of the individual hairpin structures, and elucidating how variations in their geometry influence the ejection-sweep pattern (\S\ref{sec:res-qa}) and the spatio-temporal coherence of the flow field (\S\ref{sec:res-Rii}). 
To study the phase-dependent structural characteristics of the hairpin vortex, we utilize the conditional averaging technique \citep{blackwelder1977role}. 
This technique involves selecting a flow event at a specific spatial location to condition the averaging process in time and/or space. 
The conditionally-averaged flow field is then analyzed using standard flow visualization techniques to identify the key features of the eddies involved. 
By applying this technique to the hairpin vortex, we can gain valuable insights into its structural attributes and how they vary over time.

In the past few decades, various events have been employed as triggers for the conditional averaging operation. 
For example, in the context of channel flow over aerodynamically smooth surfaces, \cite{zhou1999mechanisms} relied on an ejection event as the trigger, which generally coincides with the passage of a hairpin head through that point. 
More recently, \cite{dennis2011experimental} considered both positive cross-stream and streamwise swirl as triggers, which are indicative of hairpin heads and legs, respectively.
In flow over homogeneous vegetation canopies, \cite{watanabe2004large} used a scalar microfront associated with a sweep event. 
Shortly after, \cite{finnigan2009turbulence} noted that this choice might introduce a bias towards sweep events in the resulting structure and instead used transient peaks in the static pressure, which are associated with both ejection and sweep events.

Here, we adopt the approach first suggested by \cite{coceal2007structure}, where the local minimum streamwise velocity over a given plane was used as the trigger.
It can be shown that this approach yields similar results as the one proposed in \cite{dennis2011experimental} and that it is suitable for the identification of hairpin vortices in the ISL.
The conditional averaging procedure used in this study is based on the following operations:
\begin{enumerate}
\item Firstly, at a chosen $x_3^e$, we identify the set of locations $(x_1^e,x_2^e)$ where the instantaneous streamwise velocity is $75\%$ below its phase-averaged value. This is our ``triggering event.'' 
Such an operation is repeated for each available velocity snapshot.
\item Next, for each identified event, the fluctuating velocity field at the selected $x_3^e$ plane is shifted by $(-x_1^e,-x_2^e)$. After this operation, all identified events are located at $(x_1^\prime,x_2^\prime) = (0,0)$, where $(x_1^\prime,x_2^\prime)$ is the new (translated) coordinate system.
\item Lastly, the shifted instantaneous velocity fields are averaged over the identified events and snapshots, for each phase.
\end{enumerate}
The end result is a phase-dependent, conditionally-averaged velocity field that can be used for further analysis.
\begin{figure}
  \centerline{\includegraphics[width=\textwidth]{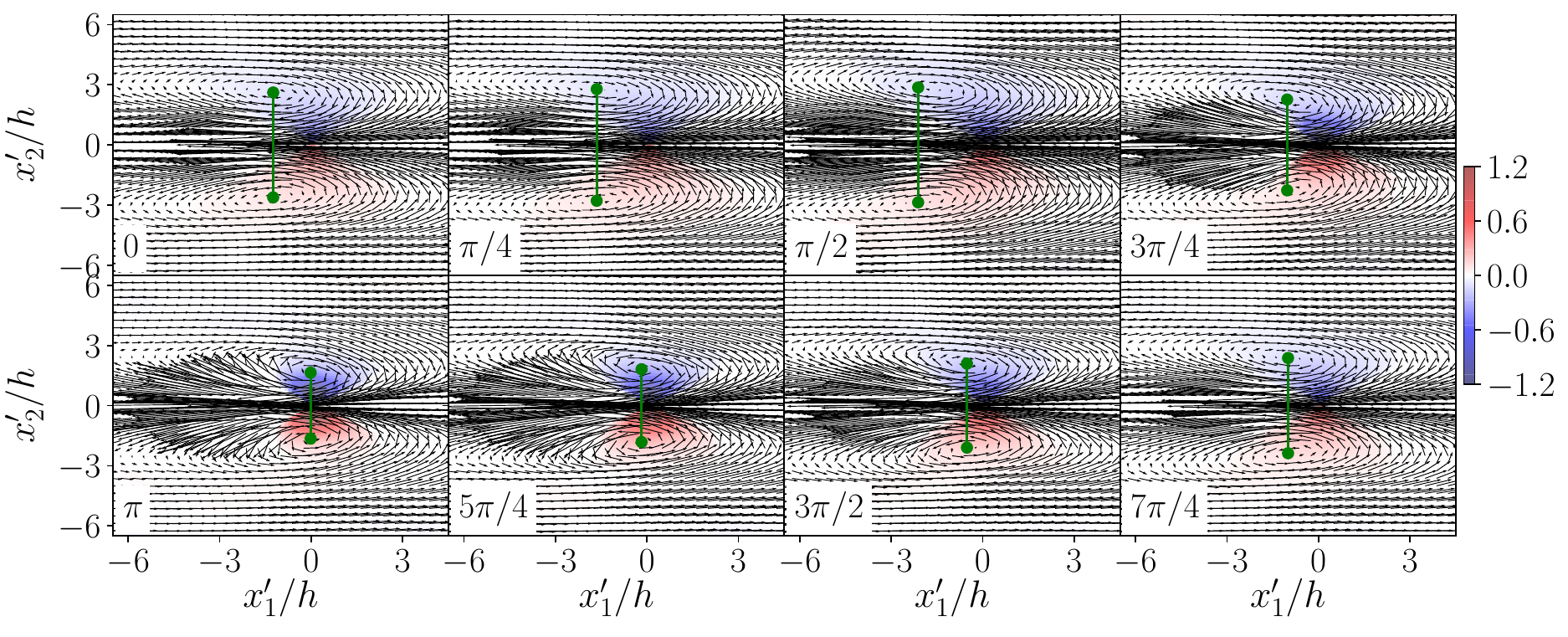}}
  \caption{Vector plot of the conditionally averaged fluctuating velocity (PP case) over the $x_3/h=2$ wall-parallel plane. The flow has been conditioned on a local minimum streamwise velocity event in the same plane. Color contours represent the wall-normal swirling strength $\lambda_{s,3}$. Green dots identify the cores of the counter-rotating vortices.}
\label{fig:cond_avg_2d_z2p0}
\end{figure}
\begin{figure}
  \centerline{\includegraphics[width=0.5\textwidth]{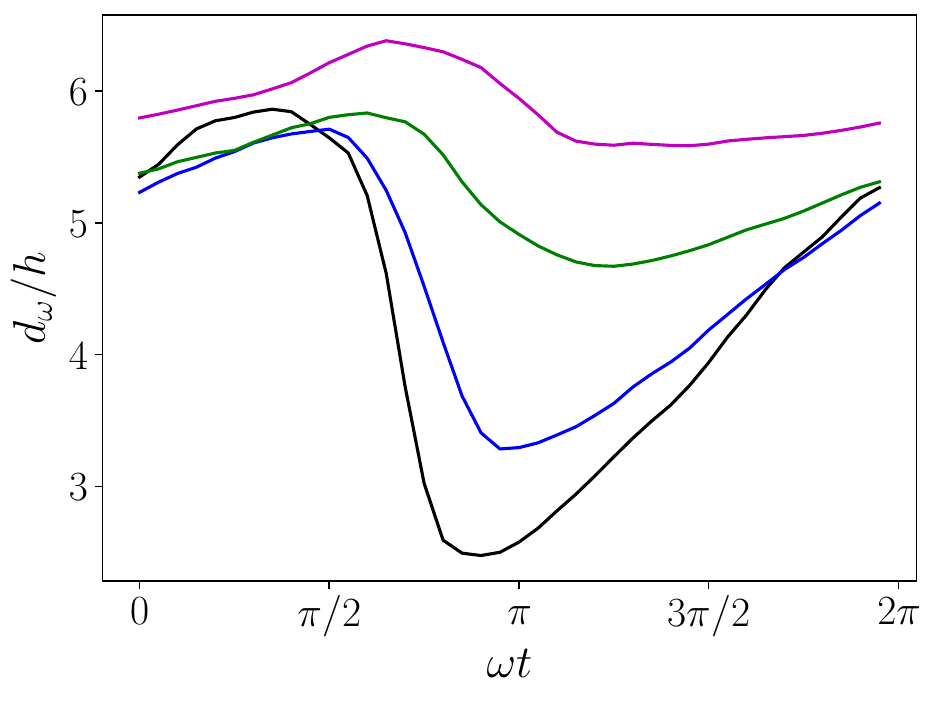}}
  \caption{Spacing between the composite vortex pair cores $d_{\omega}$, corresponding to local minimum streamwise velocity events at $x_3^e/h=1.5$ (black lines), $x_3^e/h=2$ (blue lines), $x_3^e/h=3$ (green lines) and $x_3^e/h=4$ (magenta lines).}
\label{fig:cond_avg_2d_weddy}
\end{figure}
Figure \ref{fig:cond_avg_2d_z2p0} shows a wall-parallel slice at $x_3/h=2$ of the conditionally averaged fluctuating velocity field in the same plane as the triggering event.
Counter-rotating vortices associated with a low-momentum region in between appear to be persistent features of the ISL throughout the pulsation cycle.
Vortex cores move downstream and towards each other as $ \partial \langle \overline{u}_1 \rangle/\partial x_3$ increases, and the vortices intensify. 
This behavior occurs in the normalized time interval $\omega t \in [\pi/2, \pi]$.
Instead, when $ \partial \langle \overline{u}_1 \rangle/\partial x_3$ decreases, the cores move upstream and further apart.
Such behavior provides statistical evidence of the behavior depicted in figure \ref{fig:inst_xy_z2p0}(\textit{c},\textit{d}) for the instantaneous flow field.
Note that the composite counter-rotating vortex pair in the conditionally averaged flow field is, in fact, an ensemble average of vortex pairs in the instantaneous flow field.
Thus, the spacing between the composite vortex pair cores ($d_{\omega}$) represents a suitable metric to quantify the phase-averaged widths of vortex packets in the considered flow system. 
Figure \ref{fig:cond_avg_2d_weddy} presents $d_{\omega}$ evaluated with the triggering event at $x_3^e/h=\{1.5, 2, 3, 4 \}$.
The trend in $d_{\omega}$ is similar to that observed in figure \ref{fig:ruu_xy_y0}(\textit{a}) for the first zero crossing of $\overline{R}_{11}$, which is an indicator of the streak width. 
The explanation for this behavior is that low-momentum regions are generated between the legs of the hairpins, justifying the observed linear scaling of the streak width with the cross-stream spacing of hairpin legs.

\begin{figure}
  \centerline{\includegraphics[width=\textwidth]{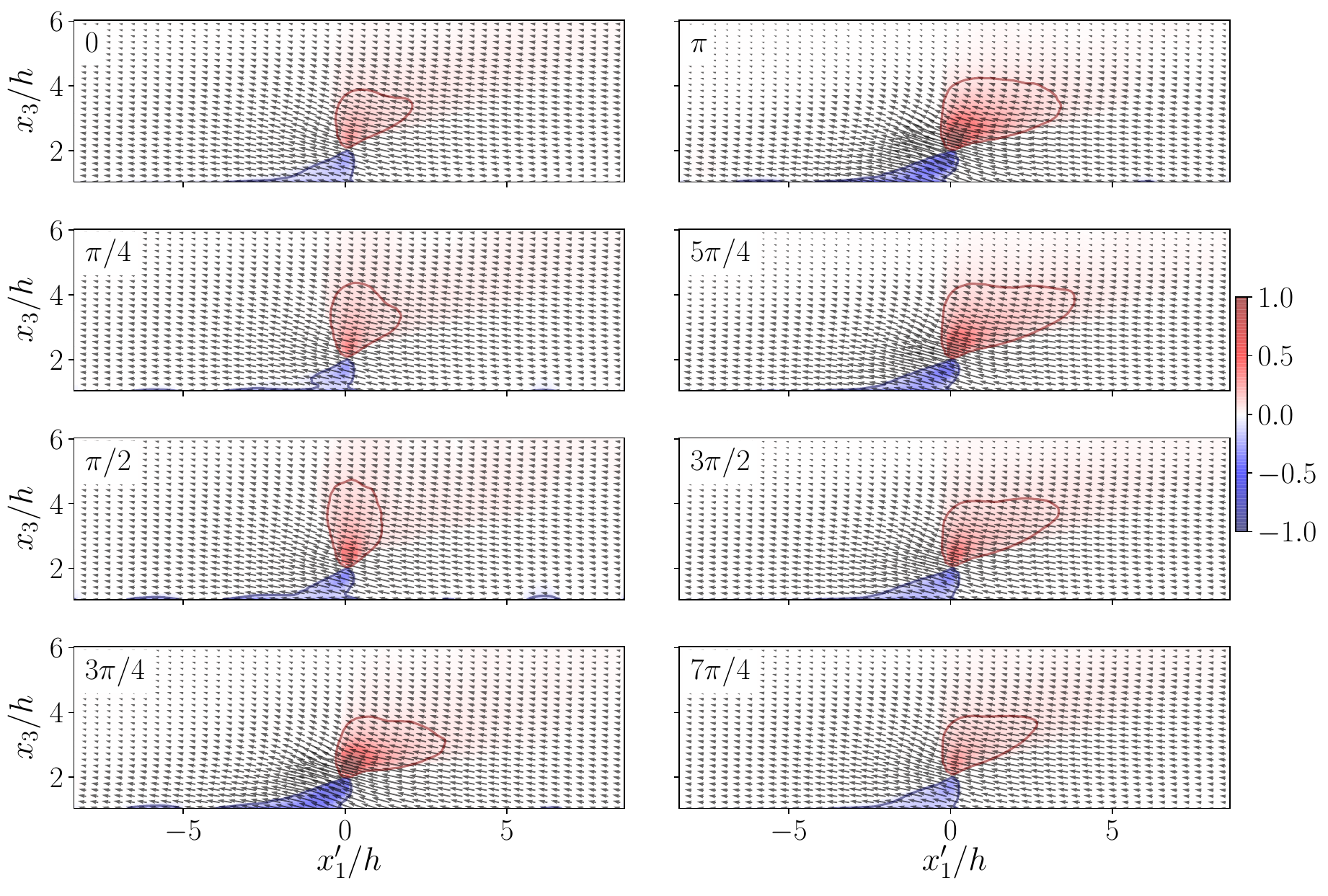}}
  \caption{Time evolution of the conditionally averaged fluctuating velocity field of the PP case in the streamwise/wall-normal plane $x_2^*/h=0$ given a local minimum streamwise velocity event at $x_3^e/h=2$. Color contours represent the cross-stream swirling strength $\lambda_{s,2}$. Red and blue lines mark the $\lambda_{s,2}=0.1$ and $\lambda_{s,2}=-0.1$ contours, respectively. }
\label{fig:cond_avg_3d_xz}
\end{figure}
\begin{figure}
  \centerline{\includegraphics[width=\textwidth]{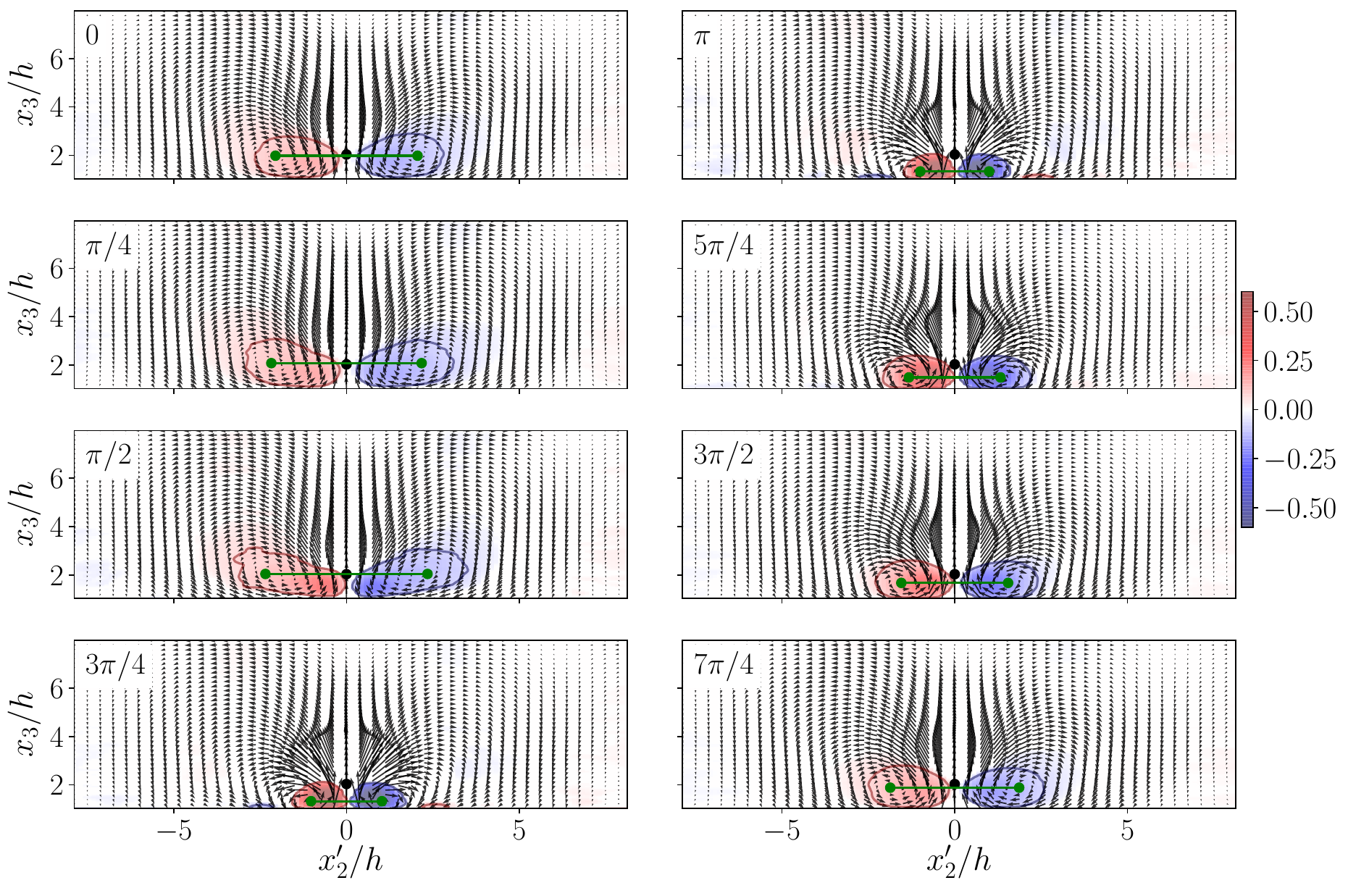}}
  \caption{Time evolution of the conditionally averaged fluctuating velocity field in figure \ref{fig:cond_avg_3d_xz} in a cross-stream/wall-normal plane $x^\prime_1=-h$. Color contours represent the streamwise swirling strength $\lambda_{s,1}$. Red and blue lines mark $\lambda_{s,1}=0.1$ and $\lambda_{s,1}=-0.1$, respectively. Green dots identify the cores of the counter-rotating vortices.}
\label{fig:cond_avg_3d_yz_upstream}
\end{figure}

Figure \ref{fig:cond_avg_3d_xz} and \ref{fig:cond_avg_3d_yz_upstream} depict a conditionally averaged fluctuating velocity field, which is obtained with a triggering event at $x_3^e/h=2$, in the $x^\prime_2=0$ plane and the $x_1^\prime=-h$ plane, respectively.
Note that the $x^\prime_2=0$ plane corresponds to the center plane, and the $x_1^\prime = -h$ cross-section is located $h$ upstream of the triggering event.
From figure \ref{fig:cond_avg_3d_xz}, a region of positive $\lambda_{s,2}$ can be identified immediately above and downstream the location of the triggering event, i.e., $(x_1^\prime,x_2^\prime,x_3^e)=(0,0,2h)$.
This $\lambda_{s,2}>0$ region can be interpreted as the head of the composite hairpin vortex \citep{adrian2000vortex,ganapathisubramani2003characteristics}. 
As $ \partial \langle \overline{u}_1 \rangle/\partial x_3$ increases, the vortex structure is deflected downstream and $\lambda_{s,2}$ increases, leading to enhanced upstream ejection events. 
This behavior is also apparent from figure \ref{fig:QA_ratio_isl}, where the overall contribution from ejection events to $\langle \overline{u_1^\prime u_3^\prime}\rangle$ increases, while the number of ejection events reduces, highlighting enhanced individual ejection events.
The deflection of the hairpin head in the downstream direction is caused by two competing factors. 
The first is the increase in $\langle \overline{u_1^\prime u_3^\prime}\rangle$, which leads to the downstream deflection. 
The second factor is the enhancement of the sweep events, which induce an upstream deflection. 
The first factor outweighs the second, thus, yielding the observed variations in the hairpin topology.

Figure \ref{fig:cond_avg_3d_yz_upstream} shows the response of hairpin legs to changing $ \partial \langle \overline{u}_1 \rangle/\partial x_3$ in a cross-stream plane at $x_1^\prime = -h$.
A pair of counter-rotating streamwise rollers are readily observed, which, as explained before, identify the legs of the composite hairpin vortex.
It also further corroborates our analysis, highlighting that the spacing between the legs reduces from $\approx 5h$ at $\omega t=\pi/2$ to $\approx 2h$ at $\omega t=\pi$. 
This also provides a justification for findings in \S\ref{sec:res-Rii} and \S\ref{sec:res-inst}.
Further, the swirling of the hairpin legs, which is quantified with $\lambda_{s,1}$ and $\lambda_{s,3}$ in the wall-normal/cross-stream and wall-parallel planes, respectively, intensifies with increasing $ \partial \langle \overline{u}_1 \rangle/\partial x_3$.
Interestingly, when $ \partial \langle \overline{u}_1 \rangle/\partial x_3$ approaches its peak value at $\omega t=\pi$, a modest albeit visible secondary streamwise roller pair is induced by the hairpin legs at $x_2^\prime=\pm3$.
This suggests that the hairpin vortex not only generates new offsprings upstream and downstream, as documented in \citep{zhou1999mechanisms,adrian2007hairpin}, but also in the cross-stream direction when it intensifies.
The intensification of hairpin legs creates counter-rotating quasi-streamwise roller pairs between the hairpin vortices adjacent to the cross-stream direction. 
These roller pairs are lifted up due to the effect of the induced velocity of one roller on the other according to the Biot–Savart law, and the downstream ends of the rollers then connect, forming new hairpin structures.

\begin{figure}
  \centerline{\includegraphics[width=\textwidth]{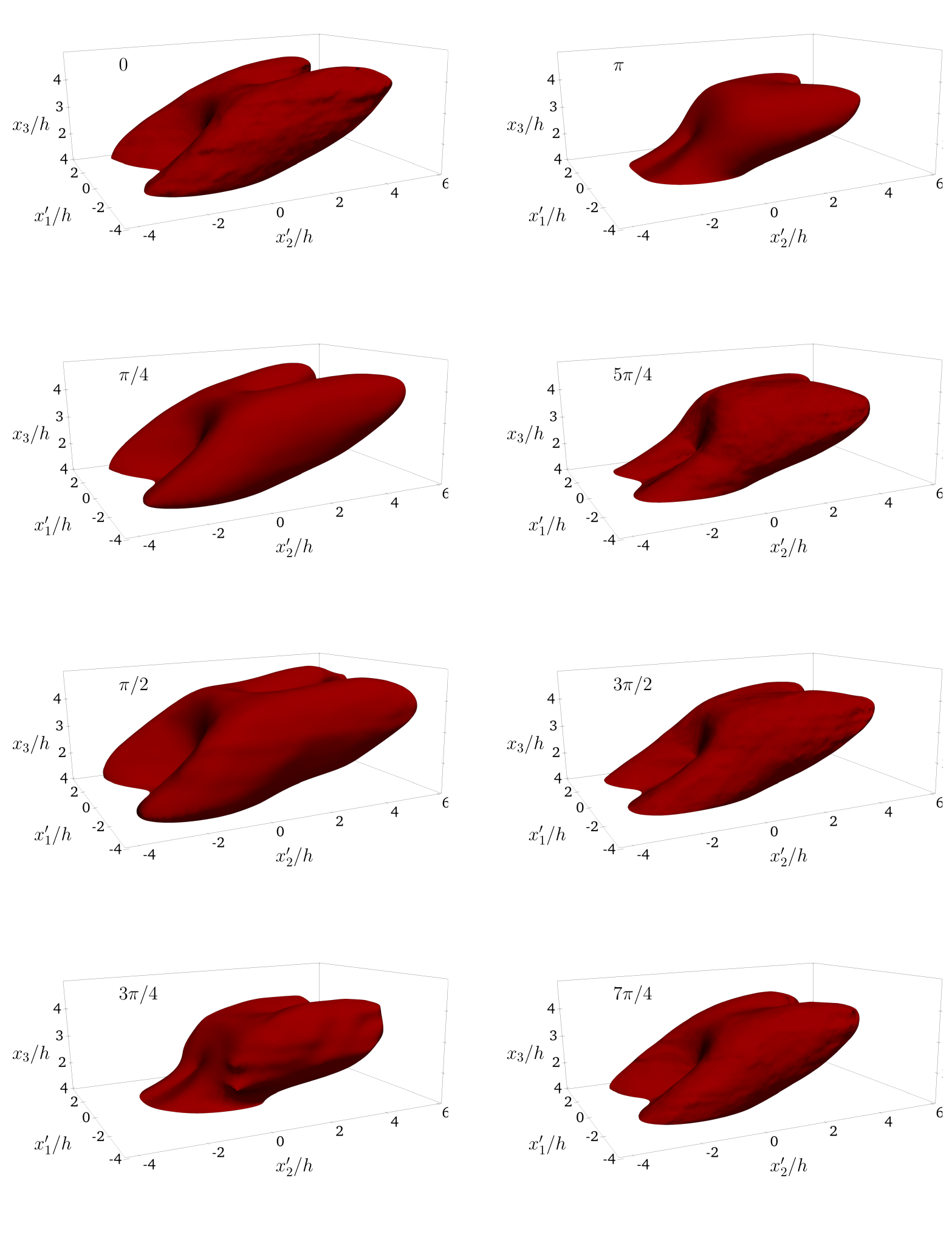}}
  \caption{Time evolution of the conditionally averaged swirling field $\lambda_s$ of the PP case given a local minimum streamwise velocity event at $x_3^e=2h$. The shown iso-surfaces are for $\lambda_s=0.1$. }
\label{fig:cond_avg_3d_swirl}
\end{figure}

A more comprehensive picture is provided by isocontours of the conditionally averaged swirling magnitude $\lambda_s = 0.1$ shown in figure \ref{fig:cond_avg_3d_swirl}. 
$\lambda_s$ is the imaginary part of the complex eigenvalue of the velocity gradient tensor \citep{zhou1999mechanisms}.
In this case, the conditionally averaged swirling field corresponds to a triggering event at $x_3^e/h=2$.
\cite{zhou1999mechanisms} pointed out that different thresholds of the iso-surface result in vortex structures of similar shapes but different sizes.
$\lambda_s=0.1$, in this case, strikes the best compromise between descriptive capabilities and surface smoothness.
Note that other vortex identification criteria, such as the Q criterion \citep{hunt1988eddies} and the $\lambda_2$ criterion \citep{jeong1995identification}, are expected to result in qualitatively similar vortex structures \citep{chakraborty2005relationships}. 

The extents of the conditional eddy in figure \ref{fig:cond_avg_3d_swirl} vary substantially from roughly $10h\times 8h \times 5h$ at relatively low $ \partial \langle \overline{u}_1 \rangle/\partial x_3$ ($\omega t=\pi/2$), to $6h\times 6h \times 3h$ at high $ \partial \langle \overline{u}_1 \rangle/\partial x_3$ ($\omega t=\pi$).
During the period of decreasing $ \partial \langle \overline{u}_1 \rangle/\partial x_3$, i.e., $0<\omega t<3/4\pi$ and $\pi<\omega t<2\pi$, the conditional eddy resembles the classic hairpin structure in the stationary case, where two hairpin legs and the hairpin head connecting the hairpin legs can be vividly observed. 
The sizes of the hairpin legs increase with decreasing $ \partial \langle \overline{u}_1 \rangle/\partial x_3$, and so does their spacing, which is in line with our prior observations based on figure \ref{fig:cond_avg_3d_yz_upstream}.
One possible physical interpretation for the change in the size of hairpin legs is that the reduction in swirling strength of the hairpin head resulting from a decrease in $ \partial \langle \overline{u}_1 \rangle/\partial x_3$ weakens the ejection between the hairpin legs, as shown in figure \ref{fig:cond_avg_3d_xz}. 
As a result, the swirling strength of the legs decreases, causing an increase in their size due to the conservation of angular momentum.
Conversely, during the period of increasing $ \partial \langle \overline{u}_1 \rangle/\partial x_3$ ($3/4\pi<\omega t<\pi$), the hairpin structure is less pronounced. 
The conditional eddy features a strengthened hairpin head, and the intensified counter-rotating hairpin legs move closer to each other and ultimately merge into a single region of non-zero swirling strength, as apparent from Figure \ref{fig:cond_avg_3d_swirl}. 
Moreover, downstream of the conditional eddy, a pair of streamwise protrusions, known as ``tongues" \citep{zhou1999mechanisms}, persist throughout the pulsatile cycle. 
According to \cite{adrian2007hairpin}, these protrusions reflect the early stage of the generation process of the downstream hairpin vortex. 
These protrusions would eventually grow into a quasi-streamwise vortex pair and later develop a child hairpin vortex downstream of the original one.

In summary, the proposed analysis reveals that the time-varying shear rate resulting from the pulsatile forcing affects the topology and swirling intensity of hairpin vortices.
As the shear rate increases (decreases), hairpin vortices tend to shrink (grow) with a corresponding enhancement (relaxation) of the swirling strength.
These variations in hairpin geometry are responsible for the observed time-varying ejection-sweep pattern (figure \ref{fig:QA_ratio_isl}).
Ejection events primarily occur between the hairpin legs, which become more widely spaced as the vortices grow and less spaced as they shrink. 
Therefore, a decrease in hairpin vortex size due to an increasing shear rate reduces the number of ejection events, while an increase in vortex size due to the decreasing shear rate leads to an increased number of ejections.
Moreover, the intensification (relaxation) of hairpin vortices at high (low) shear rates results in enhanced (attenuated) ejection events between the hairpin legs, as evidenced by figures \ref{fig:cond_avg_3d_xz} and \ref{fig:cond_avg_3d_yz_upstream}.
This enhancement and attenuation of ejection events is also corroborated by results from figure \ref{fig:QA_ratio_isl}, which indicated that high (low) shear rates decrease (increase) the number of ejection events but increase (decrease) their contribution to $\overline{u_1^\prime u_3^\prime}$.
From a flow coherence perspective, this physical process also explains the observed time evolution of $\overline{R}_{11}$ (see figures \ref{fig:ruu_xy_y0} and \ref{fig:ruu_xz_phase}), which is a statistical signature of hairpin packets.
Changes in the size of individual hairpin vortices in response to the shear rate directly influence the dimensions of hairpin packets, as the latter are composed of multiple individual hairpin structures. 

\section{Conclusions}\label{sec:conclusion}

In this study, the structure of turbulence in pulsatile flow over an array of surface-mounted cuboids was characterized and contrasted with that in stationary flow regimes. 
The objective was to elucidate the effects of non-stationarity on turbulence topology and its implications for momentum transfer.

Flow unsteadiness was observed to not significantly alter the longtime average profiles of turbulent kinetic energy and resolved Reynolds shear stress, but it marginally increased the height of the RSL.
In the context of quadrant analysis, it was found that flow unsteadiness does not noticeably alter the overall distribution within each quadrant.
However, the ejection-sweep pattern exhibited an apparent variation during the pulsation cycle.
Flow acceleration yielded a large number of ejection events within the RSL, whereas flow deceleration favored sweeps.  
In the ISL, it was shown that the ejection-sweep pattern is mainly controlled by the intrinsic and phase-averaged shear rate $ \partial \langle \overline{u}_1 \rangle/\partial x_3$ rather than by the driving pressure gradient.
Specifically, the relative contribution from ejections increases, but their frequency of occurrence decreases with increasing $ \partial \langle \overline{u}_1 \rangle/\partial x_3$. 
The aforementioned time variation in the ejection-sweep pattern was later found to stem from topological variations in the structure of ISL turbulence, as deduced from inspection of the two-point streamwise velocity correlation function and the conditionally-averaged flow field.

Specifically, the geometry of hairpin vortex packets, which are the dominant coherent structures in the ISL, was examined through the analysis of two-point velocity correlation to explore its longtime-averaged and phase-dependent characteristics.
Flow unsteadiness was found to yield relatively shorter vortex packets in a longtime average sense (up to 15\% shorter).
From a phase-averaged perspective, the three-dimensional extent of hairpin packets was found to vary during the pulsation cycle and to be primarily controlled by $ \partial \langle \overline{u}_1 \rangle/\partial x_3$, while their tilting angle remained constant throughout.
A visual examination of instantaneous structures also confirmed such behavior: the size of low-momentum regions and spacing of the hairpin legs encapsulating them were found to change with $ \partial \langle \overline{u}_1 \rangle/\partial x_3$, while the hairpin vortices remained aligned at a constant angle during the pulsation cycle.

Further insight into phase variations of instantaneous hairpin structures was later gained using conditional averaging operations, which provided compelling quantitative evidence for the behaviors previously observed. 
Specifically, the conditional averaged flow field revealed that the size and swirling intensity of the composite hairpin vortex vary considerably with $ \partial \langle \overline{u}_1 \rangle/\partial x_3$.
When $ \partial \langle \overline{u}_1 \rangle/\partial x_3$ increases to its peak value, the swirling strength of the hairpin head is intensified, yielding strengthened ejections upstream of the hairpin head and a downstream deflection of the hairpin head.
As the hairpin head intensifies, there is a corresponding increase in the intensity of the hairpin legs, coupled with a reduction in the spacing between them. This development accounts for the noted decrease in the extent of the ejection-dominated region.
In other words, individual ejections become stronger and are generated at a reduced frequency as the shear rate increases, which provides a kinematic interpretation and justification for the observed time-variability of the quadrant distribution.
Such a process, needless to say, is reversed when the shear rate decreases.

Findings from this study emphasize the significant influence that departures from statistically stationary flow conditions can have on the structure of ABL turbulence and associated processes. 
Such departures are typical in realistic ABL flows and have garnered growing attention in recent times \citep{mahrt2020non}.
While the study focuses on a particular type of non-stationarity, its results underscore the importance of accounting for this flow phenomenon in both geophysical and engineering applications. 
The modification of turbulence structures due to flow unsteadiness has a substantial effect on exchanges between the land- and ocean-atmosphere, as well as on the aerodynamic drag experienced by vehicles. 
This underlines the necessity for concerted efforts to fully characterize these modifications.
From a modeling perspective, empirical insights obtained from this study hold promise for guiding the evolution of more advanced wall-layer model formulations \citep{piomelli2008wall}. 
These models are routinely used in weather and climate forecasting, as well as in aerospace and mechanical engineering applications, facilitating the assessment of area-aggregate exchanges between solid surfaces and the adjacent fluid environment.
A recurrent shortcoming of operational wall-layer models lies in their reliance on assumptions of statistical stationarity, overlooking flow unsteadiness and state-dependent turbulence topology information \citep{Monin1954, Shamarock2008, piomelli2008wall}. This represents an important area for improvement.
Past investigations have proposed pathways to integrate turbulence topology information into wall-layer model predictions, leveraging parameters like the vortex packet inclination angle and size \citep{marusic2001experimental, marusic2010predictive}. These approaches open a fruitful avenue for assimilating the insights derived from this study into wall-layer model infrastructures.

\noindent \textbf{Declaration of Interests.} The authors report no conflict of interest. \\

\noindent \textbf{Acknowledgements.} This material is based upon work supported by, or in part by, the Army Research Laboratory and the Army Research Office under grant number W911NF-22-1-0178. 
This work used the Anvil supercomputer at Purdue University through allocation ATM180022 from the Advanced Cyberinfrastructure Coordination Ecosystem: Services \& Support (ACCESS) program, which is supported by National Science Foundation grants \#2138259, \#2138286, \#2138307, \#2137603, and \#2138296.
The authors acknowledge the Texas Advanced Computing Center (TACC) at The University of Texas at Austin for providing resources that have contributed to the research results reported within this paper.

\bibliographystyle{jfm}

\bibliography{main}

\end{document}